\documentclass[11pt]{article}
\usepackage{setspace}

\usepackage{geometry}
\usepackage{amsthm}                            
\usepackage{rotating}
\usepackage{amsmath,amssymb}
\usepackage{mathrsfs}
\usepackage{amsfonts}
\usepackage{makeidx}         
\usepackage{graphicx}        
\usepackage{multicol}        
\usepackage[bottom]{footmisc}
\usepackage{epsfig}
\usepackage{bm}
\usepackage{enumerate}
\usepackage{epstopdf}
\usepackage{multirow}
\usepackage{booktabs}
\usepackage{mathtools}
\usepackage{wrapfig}
\usepackage[pagewise, mathlines]{lineno}
\usepackage[round]{natbib}
\usepackage{caption}
\usepackage{subcaption}
\usepackage{lscape}
\usepackage[colorinlistoftodos]{todonotes}
\usepackage{float}
\usepackage{titlesec}

\mathtoolsset{showonlyrefs}
\allowdisplaybreaks
\newtheorem{theorem}{Theorem}
\newtheorem{corollary}{Corollary}
\newtheorem{lemma}{Lemma}

\newtheorem{defn}{Definition}

\newtheorem{pro}{Adaptive Procedure}

\newcommand{\pkg}[1]{{\fontseries{b}\selectfont #1}}

\titlespacing\section{0pt}{11pt plus 0pt minus 0pt}{0pt plus 2pt minus 2pt}
\titlespacing\subsection{0pt}{11pt plus 4pt minus 2pt}{0pt plus 2pt minus 2pt}
\titlespacing\subsubsection{0pt}{11pt plus 4pt minus 2pt}{0pt plus 2pt minus 2pt}

\titleformat{\section}{\normalfont\fontsize{12}{12}\bfseries}{\thesection}{1em}{}
\titleformat{\subsection}{\normalfont\fontsize{11}{11}\itshape}{\thesubsection}{1em}{}
\titleformat{\subsubsection}{\normalfont\fontsize{10}{10}\itshape}{\thesubsubsection}{1em}{}

\begin{document}

\small{

\begin{center}
\textbf{Adaptive Designs for Optimal Observed Fisher Information}
\end{center}

\begin{center}
{Adam Lane}\\
{Cincinnati Children's Hospital Medical Center}\\
{Adam.Lane@cchmc.org}\\
\vspace{0.5cm}
\end{center}

\begin{center}
{\bf Abstract}
\end{center}

\setlength{\parindent}{0pt}
Expected Fisher information can be found \textit{a priori} and as a result its inverse is the primary variance approximation used in the design of experiments. This is in contrast to the common claim that the inverse of observed Fisher information is a better approximation of the variance of the maximum likelihood estimator. Observed Fisher information cannot be known \textit{a priori}; however, if an experiment is conducted sequentially, in a series of runs, the observed Fisher information from previous runs is known. In the current work, two adaptive designs are proposed that use the observed Fisher information from previous runs to inform the design of future runs.
}

\section{Introduction} \label{sec:intro}


The accuracy of an estimate is influenced by several factors including the quality of the estimator, the efficiency of the design, and even luck in the form of \textit{experimental error} - the error associated with a specific set of observed data. Before the data is observed we, naturally, ignore the experimental error and assess the quality of an estimator by averaging over all possible outcomes. In optimal design, the quality of a design is defined as a function of the expected Fisher information (EFI). Inverse EFI is a measure of the average accuracy of the maximum likelihood estimator, the estimator considered in this paper. Use of the optimal design with respect to the maximum likelihood estimator addresses the first two factors in the accuracy of an estimate. The remaining factor, experimental error, is an attribute of the observed data. This discussion reveals a limitation in the optimal design of an experiments. Specifically, an optimal design maximizes the efficiency of the \textit{estimator}; however, it does not guarantee the efficiency of the \textit{estimate}. The main objective of this paper is to develop a design that adapts as the data are observed to increase the efficiency of the maximum likelihood estimate (MLE). Since, we propose adapting the design as data are observed our proposed methods require observations can be collected sequentially, in a series of runs. In a sequential experiment the observed data from the previous runs is available to incorporate into the design of future runs.

The dependence of the accuracy of an estimate on the sample has been studied rigourously from various perspectives including, conditional inference, expected loss and the Bayesian framework. The accuracy of an estimate is tied to the information contained in the sample. We consider the information in the sample as a synonym for the accuracy and efficiency of the estimate.

Conditional inference is the most widely researched prespective. Briefly, suppose a minimal sufficient statistic is comprised of the MLE, $\hat{\theta}$, and an ancillary random variable, $a$. A random variable is \textit{ancillary} if its distribution is independent of the model parameters. An ancillary random variable is considered \textit{maximally} ancillary if it is a function of the minimal sufficient statistic. \citet*{Fish:Theo:1925,Fish:TwoN:1934} showed that the conditional distribution of $\hat{\theta}$ given $a$ contains all the information in the sample. Therefore, the accuracy of the MLE is best described by the conditional variance, Var[$\hat{\theta}|a]$. Fisher's original statements have been supported and expanded in \citet*{Cox:Some:1958}, \citet*{Berg:Wolp:TheL:1984}, \citet*{Fras:Anci:2004}, \citet*{Ghos:Reid:Fras:Anci:2010}, and others.

Summarizing a statement by Cox in an interview with \citet*{Reid:Acon:1994}: the argument for conditional inference is overwhelming; however, to convert this to practice is a severe challenge. A solution is to rely on approximation. In a landmark paper \citet*{Efro:Hink:Asse:1978} show, in the single parameter translation family, that the difference between Var[$\hat{\theta}|a]$ and the inverse of observed Fisher information (OFI), sometimes referred to as simply observed information, is of the order $O_{p}(n^{-1})$. Conversely, the difference between the Var[$\hat{\theta}|a]$ and the inverse of EFI has order $O_{p}(n^{-1/2})$. The generalization of this result remains an open question; however, many empirical studies have demonstrated the superiority of OFI. \citet*{Efro:Hink:Asse:1978} and \citet*{McCu:Loca:1984} examine OFI in more general single parameter distributions. \citet*{Barn:Sore:ARev:1994} consider multivariate generalizations. Observed information has proved useful in many areas of statistical research including the EM algorithm [\citet*{Loui:Find:1982}], generalized linear models [\citet*{Firt:Bias:1993}], semi-parametric models [\citet*{Murp:Van:Obse:1999}], hidden markov models [\citet*{Lyst:Hugh:Exac:2002}] and likelihood theory [\citet*{Reid:Asym:2003}] to name a few. \citet*{LinFlouRose2019} found that normalizing parameter estimates using OFI aids in the analysis of adaptive experiments. Based on these insights the present author shares the opinion of \citet*{Barn:Sore:ARev:1994} in that we ``generally consider observed information as a standard with which the other measures are to be compared."  

From a perspective of expected loss, \citet*{Lind:Li:Onse:1997} demonstrate the optimality of OFI with respect to the realized square error $n(\hat{\theta} - \theta)^{2}$. Specifically, in a general setting, they show that the loss estimator $T(X)$ that minimizes of the asymptotic mean square error criterion, i.e.,
\begin{align}
\min E[\{n(\hat{\theta} - \theta)^{2} - T(X) \}^{2}],
\end{align}
is the inverse of OFI, up to a $O(n^{-3/2})$ error term. In this way OFI is optimal with respect a class of estimators that includes the inverse EFI, bootstrap estimators and jackknife estimators, which all have error $O(n^{-1})$.

From the Bayesian perspective, the posterior distribution is conditional on the observed data. It has been observed [see \citet{Berg:Stat:1985}] that the posterior covariance matrix is more closely approximated by the inverse of OFI than the inverse of EFI.

From all three perspectives, the consensus is that OFI best represents the information in the sample. As such incorporating OFI into the design will increase the efficiency of the estimate. In this paper two adaptive designs are proposed that use the well developed tools of optimal design to ``minimize" inverse OFI in a general setting. This is analogous to the objective in optimal design of minimizing the inverse EFI. In general EFI is a matrix and therefore the minimum of its inverse with respect to all objectives does not exist. Instead a criterion is selected according to the primary objective of the experiment. For example, if it is desired to minimize the volume of the confidence ellipsoid of the parameter estimates, the $D$-optimal criterion (the determinant of the inverse of EFI) is applied. For an in depth description of optimal design and optimality criteria see \citet*{Atki:Done:Tobi:opti:2007}.


Adaptive designs have been considered in the literature in the context of ``locally" optimal designs. A design is \textit{locally optimal} if the EFI depends on the model parameters; since, such designs are optimal only in a neighborhood of the true parameters. One common approach is to evaluate the locally optimal design based on a fixed \textit{a priori} guess of the true parameters. We will refer to this as the fixed locally optimal design (FLOD). Alternatively, adaptive designs have been proposed as early as \citet*{Box:Hunt:Sequ:1965}. Such designs are often referred to as \textit{adaptive optimal designs} (AOD) in that they evaluate EFI at the MLE calculated from the data of all previous runs to estimate the optimal design for the current run. \citet*{Drag:Fedo:Adap:2005}, \citet*{Drag:Husua:Padm:adap:2007}, \citet*{Drag:Fedo:Wu:Adap:2008}, \citet*{Lane:Yao:Flou:Info:2014} and \citet*{Kim:Flou:2014} examine AODs.


The efficiency of a design is traditionally defined in terms of EFI. This is inadequate to describe the efficiency of the estimate, which is a function of the observed data. In Section \ref{sec:bench} a measure of observed efficiency is proposed. This definition is used to motivate and develop designs to minimize the inverse of OFI. 

The remainder of the paper is structured as follows. In Section \ref{sec:Adapt_Proc} we propose a local adaptive design based on the OFI evaluated at the same fixed \textit{a priori} guess of the parameters for all runs. This design is analogous to the FLOD. There are a few reasons we consider a local approach. \citet*{Dett:Born:OnTh:2013} and \citet*{Lane:Yao:Flou:Info:2014} both found heuristically that for finite sample sizes the use of the FLOD often outperforms the AOD. The local approach also illustrates the benefits of designs based on OFI without being confounded by a procedure where the MLE is used to update the initial guess. Further, we show that the proposed local adaptive design has maximum local observed efficiency up to a $O_{p}(n^{-1})$ error. For comparison, the FLOD has maximum local observed efficiency with a $O_{p}(n^{-1/2})$ error. A second adaptive procedure is proposed in Section \ref{subsec:adapt_update} that incorporates OFI and updates the initial guess using the MLE based on previous runs. This procedure is analogous to existing AOD approaches. 

It is important to highlight a difference between EFI and OFI in terms of the local dependence on $\theta$. It is possible for EFI to be independent of $\theta$; but for OFI to be a function of $\theta$. Such an example is given in Section \ref{sec:GHR} for a generalized linear model (GLM) with gamma distributed responses. In such cases the optimal design is \textit{global} since it maximizes the efficiency over the entire parameter space. However, it is only efficient in terms of the EFI. The procedures proposed in this paper can still be used in to improve designs. It is shown, heuristically in a simulation study, that the proposed procedures significantly increases the efficiency, both in terms of the observed efficiency and the unconditional variance of the parameter estimates, compared to the FLOD and the AOD. This indicates that the designs proposed in this paper have a broader application than AODs.

\section{Model, Information and Design} \label{sec:elem}

We assume a response $y$ has density $f(y|\eta)$, where $\eta$ is a scale parameter within $H$. The parameter $\eta = \eta_{\theta}(x)=\theta^{T}f_{x}(x)$ is a linear function of the model parameters $\theta$; where $\theta$ is a $p$ dimensional vector within the parameter space $\Theta$; $f_{x}(x)$ is a function of the design points $x\in\mathcal{X}$; and $\mathcal{X}$, the design region, is a compact subset of $\mathbb{R}^{s}$. The dependence of the function $\eta$ on $x$ and $\theta$ will be omitted when the meaning is clear. It is assumed that $y_{1}|\eta, y_{2}|\eta, \ldots$ are independent for all $\eta \in H$. If $f(y|\eta)$ is in the exponential family then it is a generalized linear model; however, we do not restrict the distribution in this way.

From a design perspective we assume independent replications are possible from the underlying model and that the value of the design points, $x$, at which the replications will be observed are under the purview of the experimenter. We define an \textit{exact design} as a collection of $d$ unique design points, $x_{1}, \ldots, x_{d}$, with corresponding allocation weights, $w_{i} = n_{i}/n$ and denote it as
\begin{align}
\xi_{n} = \left\{\begin{array}{cccc} x_{1} & \ldots & x_{d} \\ w_{1} & \ldots & w_{d} \end{array}\right\},
\end{align}
where $n = \sum n_{i}$ is the total sample size.

There are two sources of information that are important to distinguish for the designs proposed in this paper. The first is the elemental information, i.e., the information from a singe subject with respect to $\eta$ collected at design point $x$. Denote the log likelihood as $l_{\eta}(x,y) = \log f(y|\eta)$. We define \textit{observed elemental information} as
\begin{align}
I_{\eta}(x,y) &= - \frac{\partial^{2}}{\partial \eta^{2}} l_{\eta}(x,y)
\end{align}
and, assuming derivatives and integrals are exchangeable, the \textit{expected elemental information} as
\begin{align}
\mu_{\eta}(x) &= E[I_{\eta}(x,y)].
\end{align}
\citet{Atki:Fedo:Herz:Elem:2014} demonstrate that the expected elemental information is crucial in finding solutions to many optimal design problems. Similarly, in Section \ref{sec:Adapt_Proc} we demonstrate that the ratio of observed to expected elemental information is crucial in motivating and finding adaptive designs to maximize OFI. We denote the sum, over all observations at the $i$th design point, of the ratio of observed elemental information to expected elemental information as
\begin{align}
q_{\eta}(\mathcal{D}_{n_{i}}) = \sum_{j=1}^{n_{i}} I_{\eta}(x_{i},y_{ij}) / \mu_{\eta}(x_{i}).
\end{align}

The second source of information is with respect to $\theta$. This source describes the accuracy of the MLE. First we collect some necessary notation. Let $\bm{y_{i}}=(y_{i1},\ldots,y_{in_{i}})^{T}$ be the $n_{i}$ responses observed at the support point $x_{i}$; $\mathcal{D}_{n_{i}}  = (x_{i},\bm{y_{i}})$ be the available data at $x_{i}$; and $\mathcal{D}_{n} = (\mathcal{D}_{n_{1}},\ldots,\mathcal{D}_{n_{d}})$ be all available data collected from an experiment of size $n$.

The MLE is defined as
\begin{align}
\hat{\theta}_{n} = \hat{\theta}_{n}(\mathscr{D}_{n}) = \arg\max_{\theta\in\Theta} \prod_{i=1}^{d} \prod_{j=1}^{n_{i}} f\{y_{ij}|\eta_{\theta}(x_{i})\}.
\end{align}
The MLE is a function of the observed data and is therefore a function of the given design $\xi_{n}$. To specify from which design the MLE was obtained, we may write $\hat{\theta}_{n} =\hat{\theta}(\xi_{n})$. Throughout we assume standard regularity conditions are satisfied [see for example \citet*{Ferg:ACou:1996} chapters 17 and 18] to ensure that the MLE is strongly consistent and asymptotically normal with asymptotic variance-covariance matrix equal to the inverse of the normalized EFI. It is well known that the normalized EFI is
\begin{align} \label{eq:mu}
M_{\theta}(\xi_{n}) = \sum_{i=1}^{d} w_{i} \mu_{\eta}(x_{i}) f_{x}(x_{i}) f_{x}^{T}(x_{i}).
\end{align}
Note, throughout we use several shorthand notations. These are collected in Table \ref{tab:notatoin} for reference.
\begin{table}
\centering
\scriptsize
\begin{tabular}{ ccc }
\noalign{\smallskip}\hline\noalign{\smallskip}
Shorthand & Notation & Brief Description  \\
\noalign{\smallskip}\hline\noalign{\smallskip}
MLE & $\hat{\theta}_{n}$ & Maximum Likelihood Estimate \\
EFI & $nM(\xi_{n})$ & Expected Fisher Information \\
OFI & $J_{\theta}(\mathscr{D}_{n})$ & Observed Fisher Information \\
FLOD & $\xi_{\Psi}^{*}(\theta)$ & Fixed Local Optimal Design \\
AOD & $\overline{\xi}_{\Psi}(\theta)$ & Adaptive Optimal Design \\
LOAD & $\tilde{\xi}_{\Psi}$ & Local Observed Information Adaptive Design \\
MOAD & $\hat{\xi}_{\Psi}$ & Maximum Likelihood Estimated Observed Information Adaptive Design \\
$\Psi_{\rm{eff}}^{\theta}[\tau_{\eta}]$ & $\frac{\Psi[M_{\theta}\{\xi_{\Psi}^{*}(\theta)\}]}{\Psi[M_{\theta}\{\tau_{\eta}(\mathcal{D}_{n})\}]}$  & Local Observed $\Psi$ efficiency \\
$\Psi_{\rm{eff}}^{\hat{\theta}_{n}}[\tau_{\hat{\eta}}]$ & $\left[\frac{\Psi[M_{\theta}\{\xi_{\Psi}^{*}(\theta)\}]}{\Psi[M_{\theta}\{\tau_{\eta}(\mathcal{D}_{n})\}]}\right]_{\theta = \hat{\theta}_{n}}$  & Observed $\Psi$ efficiency at the MLE \\
$\mbox{Rel-Eff}_{\Psi}$ & $\frac{\Psi\{\mbox{Var}[\hat{\theta}(\xi_{1})]^{-1}\}}{\Psi\{\mbox{Var}[\hat{\theta}(\xi_{2})]^{-1}\}}$ & Relative Efficiency \\
\noalign{\smallskip}\hline\noalign{\smallskip}
\end{tabular}
\caption{Shorthand notation for commonly used definitions and formulas.} \label{tab:notatoin}
\end{table}

Let $Q_{\eta}(\mathcal{D}_{n}) = \sum_{i=1}^{d} q_{\eta}(\mathcal{D}_{n_{i}})$ be the total of the ratios of observed to expected elemental information across all design points. Then, assuming $Q_{\eta}(\mathcal{D}_{n}) \ne 0$, let
\begin{align}
\omega_{\eta}(\mathcal{D}_{n_{i}}) = \frac{1}{Q_{\eta}(\mathcal{D}_{n})} q_{\eta}(\mathcal{D}_{n_{i}})
\end{align}
for $i=1,\ldots,d$; and
\begin{align} \label{eq:tau}
\tau_{\eta}(\mathcal{D}_{n}) = \left\{\begin{array}{cccc} x_{1} & \ldots & x_{d} \\ \omega_{\eta}(\mathcal{D}_{n_{1}}) &  \ldots & \omega_{\eta}(\mathcal{D}_{n_{d}}) \end{array}\right\}.
\end{align}
Then for any observed data set, $\mathcal{D}_{n}$, we can write the normalized OFI as
\begin{align}
\frac{1}{n}J_{\theta}(\mathcal{D}_{n}) 
&= \frac{1}{n}\sum_{i=1}^{d} \sum_{j=1}^{n_{i}} I_{\eta}(x_{i},y_{ij})f_{x}(x_{i}) f_{x}^{T}(x_{i}) \\
&= \frac{1}{n}Q_{\eta}(\mathcal{D}_{n})M_{\theta}\{\tau_{\eta}(\mathcal{D}_{n})\} \label{eq:J}.
\end{align}
This representation shows that the normalized OFI is proportional to the normalized EFI evaluated at $\tau_{\eta}$.

In Section \ref{sec:bench} this representation is used to derive a definition of observed efficiency which is used in the derivation of the proposed local adaptive designs in Section \ref{sec:Adapt_Proc}.

\subsection{Fixed Optimal Design} \label{subsec:trad_opt_design}

The traditional objective in optimal design is to find the design that minimizes EFI, with respect to a concave criterion function, denoted $\Psi(\cdot)$. In practice, the optimality criterion, $\Psi$, is selected according to the primary objective of the experiment. Let $\mathcal{S}_{+}^{p}$ be the set of $p\times p$ symmetric positive semi-definite matricies. Then $\Psi$ is a mapping from $\mathcal{S}_{+}^{p}$ to $R^{+}\coloneqq[0,\infty)$. The design problem is the to find the design $\xi_{\Psi}^{*}(\theta)$ which is defined as
\begin{align} \label{eq:Opt}
\xi_{\Psi}^{*}(\theta) = \arg \min_{\xi \in \Xi} \Psi\{M_{\theta}(\xi)\},
\end{align}
where $\Xi$ represents the set of all permissible designs. When EFI depends on the parameter $\theta$ this is the fixed local optimal design (FLOD); local since it is a function of $\theta$; and fixed in the sense that it is determined before the experiment (not adaptive).

Two classical approaches to the optimal design problem are defined by the set over which the maximization in equation \eqref{eq:Opt} takes place. For \textit{exact optimal design}, the maximization is over the set of all possible exact designs $\Xi_{n}$. To find exact optimal designs search algorithms are commonly employed. Classical examples of search algorithms are found in \citet*{Fedo:Theo:1972}, \citet*{Mitc:AnAl:1974} and \citet*{Atki:Done:Tobi:opti:2007}. The search for exact optimal designs can be difficult and computationally expensive. \textit{Continuous optimal design} relaxes the restriction that $nw_{i}$ be an integer to the restriction that $0 \le w_{i} \le 1$ and $\sum_{i} w_{i} = 1$ for $i=1,\ldots,d$. We denote a continuous design by $\xi$ and note that it is a member of the set of all possible continuous designs denoted $\Xi_{\Delta}$. Note $\Xi_{n}$ is a subset of $\Xi_{\Delta}$. Once again the common approach is to solve equation \eqref{eq:Opt} using a search algorithm. Classical examples of algorithms for continuous designs are found in \citet*{Wynn:TheS:1970} and \citet*{Fedo:Theo:1972}. Of course, all designs implemented in an experiment are exact. Rounding methods to convert a continuous design into an exact design are considered in \citet*{Puke:Ried:Effi:1992}.  There exists a rich literature in both the exact and continuous optimal design problems.

In principle $\Psi$ can be any arbitrary convex optimality criterion. For illustrative purposes we consider the $D$ and $A$ optimality criteria, where $\Psi(M)$ is defined as $|M^{-1}|^{1/p}$ and $\mathrm{Tr}(M^{-1})\}$, respectively.

\section{Local Observed Efficiency} \label{sec:bench}
The objective of an optimal design is to maximize the efficiency of an experiment. In this section we begin by deriving the definition of efficiency most commonly used in optimal design. We then propose an analogous measure of observed efficiency that is more consistent with measuring the efficiency of the design as it relates to the observed data.

The relative efficiency of any two designs, say $\xi_{1}$ and $\xi_{2}$, can be assessed, with respect to the criterion $\Psi$, as
\begin{align}
\mbox{Rel-Eff}_{\Psi}(\xi_{1},\xi_{2}) = \frac{\Psi\{\mbox{Var}[\hat{\theta}(\xi_{1})]^{-1}\}}{\Psi\{\mbox{Var}[\hat{\theta}(\xi_{2})]^{-1}\}}.
\end{align}
If $\mbox{Rel-Eff}_{\Psi}(\xi_{1},\xi_{2}) \le 1$, then $\xi_{1}$ is the more efficient design. In many cases the variance is not tractable and is  replaced with its asymptotic approximation, $\{nM_{\theta}(\xi)\}^{-1}$. This replacement has an error of $O(n^{-1})$, in general [see \citet{McCu:Tens:1987} chapter 7]. From a design perspective the primary interest is often to define the efficiency of an arbitrary exact design relative to the continuous optimal design. This yields the definition of local $\Psi$-efficiency most commonly used in optimal design
\begin{align}
\Psi_{\rm{eff}}^{\theta}(\xi) &= \frac{\Psi[M_{\theta}\{\xi_{\Psi}^{*}(\theta)\}]}{\Psi\{M_{\theta}(\xi)\}}. \label{eq:Geff}
\end{align}
From the definition of the optimal design, $\Psi[M_{\theta}\{\xi_{\Psi}^{*}(\theta)\}]$ is a lower bound and thus $\Psi_{\rm{eff}}^{\theta}(\xi)\le1$ for all $\xi\in\Xi_{\Delta}$.

Returning to the language of the introduction, both of the two preceding measures of efficiency relate to the efficiency of the maximum likelihood \textit{estimator}. They are deficient at precisely describing the efficiency of the maximum likelihood \textit{estimate} due to its dependence on the observed data.  From the conditional perspective of Fisher, given in the introduction, a more relevant measure of efficiency is what we define as the \textit{conditional relative efficiency} of $\xi_{2}$ relative to $\xi_{1}$, namely,
\begin{align}
\mbox{Cond-Rel-Eff}_{\Psi}(\xi_{1},\xi_{2}) = \frac{\Psi(\mbox{Var}[\hat{\theta}(\xi_{1})|a]^{-1})}{\Psi(\mbox{Var}[\hat{\theta}(\xi_{2})|a]^{-1})}.
\end{align}
In this definition, $a$ is a maximal ancillary random variable. There three problems with the practical value of this definition related to the design of experiments.
\begin{enumerate}
\item{The conditional variance of the MLE has no general solution.}
\item{There is no known lower bound to reduce the problem to an assessment of one design against the best possible design. }
\item{The maximal ancillary random variable, $a$, is a function of the observed data, therefore, the relative efficiency of any two designs is data dependent. This increases the design problem significantly, since in evaluating the efficiency of any two arbitrary designs we may find that design $\xi_{1}$ may be more efficient for some values of $a$ yet $\xi_{2}$ could be more efficient for other values. }
\end{enumerate}

Motivated by the result of \citet*{Efro:Hink:Asse:1978} we solve problem 1 by replacing $\mbox{Var}[\hat{\theta}(\xi)|a]$ with the approximation $J_{\theta}^{-1}$. Implementing this substitution requires that we extend the previous definition of $\Psi$ to nonpositve definite matrices. When $J_{\theta}$ is evaluated at an arbitrary $\theta$ is not necessarily positive definite. The optimality criteria $\Psi$ is not defined on the set of non-positive definite matrices. A convention is to define $\Psi(M)$ to be equal to zero when $M$ is singular. Here we extend this convention by defining $\Psi(M)$ to be equal to zero if $M$ is negative definite. When $J_{\theta}$ is evaluated at $\hat{\theta}$ it is guaranteed to be positive semi-definite [see \citet*{Barn:Sore:ARev:1994}]. With respect to this substitution we present a definition of \textit{observed efficiency} that is analogous to the definition used in optimal design.
\begin{defn}  \label{def:Dobs_eff} For any observed data, $\mathcal{D}_{n}$, the local observed $\Psi$-efficiency is
\begin{align} \label{eq:Dobs_eff}
\Psi_{{\rm{obs-eff}}}^{\theta} (\mathcal{D}_{n}) &= \frac{\Psi[Q_{\eta}(\mathcal{D}_{n}) M_{\theta}\{\xi_{\Psi}^{*}(\theta)\}]}{\Psi\{J_{\theta}(\mathcal{D}_{n})\}} = \frac{\Psi[M_{\theta}\{\xi_{\Psi}^{*}(\theta)\}]}{\Psi[M_{\theta}\{\tau_{\eta}(\mathcal{D}_{n})\}]} \\
&=\Psi_{\rm{eff}}^{\theta}\{\tau_{\eta}(\mathcal{D}_{n})\}.
\end{align}
\end{defn}
The local observed $\Psi$-efficiency is simply the $\Psi$-efficiency with respect to the ``design'' $\tau_{\eta}$ given in \eqref{eq:tau}.  Throughout we use $\Psi_{\rm{eff}}^{\theta}(\tau_{\eta})$ to represent the local observed $\Psi$-efficiency as shorthand. The following Lemma establishes a lower bound on the OFI with respect to $\Psi$.
\begin{lemma} \label{thm:Q}
If $I_{\eta}:R^{s+1}\rightarrow R^{+}$ and $\theta\in\Theta$ then
\begin{align} \label{eq:obs_bound}
\Psi[M_{\theta}\{\tau_{\eta}(\mathcal{D}_{n})\}] \ge \Psi[M_{\theta}\{\xi_{\Psi}^{*}(\theta)\}]
\end{align}
for any observed data, where $\mathcal{D}_{n}$where  $\xi_{\Psi}^{*}(\theta)$ be the optimal design on the approximate design space $\Xi_{\Delta}$.
\end{lemma}
The proof follows directly from the definition of the optimal design given in equation \eqref{eq:Opt}.  Under the conditions of Lemma \ref{thm:Q}, $\Psi_{{\rm{eff}}}^{\theta} \le 1$. This solves problem 2 since it is now possible to assess the observed efficiency of a single design compared the lower bound.

From Definition \ref{def:Dobs_eff} $\Psi_{\rm{eff}}^{\theta}(\tau_{\eta}) = 1$ if $\tau_{\eta}$ is equal to $\xi_{\Psi}^{*}(\theta)$. From \eqref{eq:tau} this occurs if $\omega_{\eta}(\mathcal{D}_{n_{i}})$ is equal to $w_{i}^{*}$ for all $i=1,\ldots,d$. Now we can better explained how problem 3 can be addressed in a sequential experiment. It is desirable for the absolute differences between $|\omega_{\eta}(\mathcal{D}_{n_{i}}) - w_{i}^{*}|$ to be minimized. In a sequential experiment these absolute differences can be examined after each run and the future runs can be assigned to design points where large differences have accrued. This provides intuition for the design proposed in the next section where the theory of optimal design is used to develop such an adaptive procedure.

One final remark. The condition in Lemma \ref{thm:Q} that $I_{\eta}:R^{s+1}\rightarrow R^{+}$ is moderately restrictive. This assumption guarantees the inequality of the lemma holds; however, in practice the result of the lemma usually holds even in cases when this condition does not. Further, this condition is not required for the designs proposed in the next two sections.

\section{Locally Adaptive Design} \label{sec:Adapt_Proc}

In this section, the first of two adaptive designs incorporating OFI is developed. This design is local in that it is implemented at a fixed initial guess of $\theta$ for all runs. The local approach reduces design complexity, can be implemented in cases where OFI does not depend on $\theta$ and has been found to be robust in the context of AODs  [see \citet*{Dett:Born:OnTh:2013} and \citet*{Lane:Yao:Flou:Info:2014}].  Its implementation requires that the FLOD can be found, either exactly or approximately.

\subsection{Sequential Experiment}

Consider an experiment with $J$ sequential runs. Each run has fixed number, $m_{j}$, of independent observations for $j=1,\ldots,J$. Denote the total sample size up to and including the $j$th run as $M_{j} = \sum_{k=1}^{j} m_{k}$ and note $n = M_{J}$. To frame the problem similarly to fixed optimal design, we define the design of the $j$th run as
\begin{align}
\lambda_{j} = \left\{\begin{array}{ccc} x_{1j} & \ldots & x_{d_{j}j} \\ w_{1j} & \ldots & w_{d_{j}j} \end{array}\right\},
\end{align}
where $x_{ij}\in\mathcal{X}$ are the design points and $w_{ij}$ are the corresponding allocation weights, $i=1,\ldots,d_{j}$, at run $j$, with $0\le w_{ij}\le 1$ and $\sum_{i=1}^{d_{j}}w_{ij}=1$. If $m_{ij} = m_{j}w_{ij}$ are integers for all $i=1,\dots,d_{j}$ then $\lambda_{j}$ is an exact design.

Let $\mathcal{D}_{M_{j}}$ represent all available data collected from the first $j$ runs. Now we define a quantity that represents the EFI of the $j$th run, given the design of the current run, $\lambda_{j}$, and the observed data from the previous $j-1$ runs as
\begin{align}
K_{\theta}(\lambda_{j},\mathcal{D}_{M_{j-1}}) &= E[J_{\theta}(\mathcal{D}_{M_{j}})|\lambda_{j},\mathcal{D}_{M_{j-1}}] \\
&= m_{j} M_{\theta}(\lambda_{j}) + J_{\theta}(\mathcal{D}_{M_{j-1}}) \\
&= m_{j} M_{\theta}(\lambda_{j}) + Q_{\eta}(\mathcal{D}_{M_{j-1}})M_{\theta}\{\tau_{\eta}(\mathcal{D}_{M_{j-1}})\} \label{eq:K}.
\end{align}
An intuitive objective is to find the design that minimizes $K_{\theta}$ with respect to some criterion function. This is an optimal design problem and can be stated precisely as finding
\begin{align}
\lambda_{j}^{*}(\theta) = \arg\min_{\lambda_{j} \in \Xi_{\Delta}} \Psi\{K_{\theta}(\lambda_{j},\mathcal{D}_{M_{j-1}})\}. \label{eq:DK_op}
\end{align}
Solving equation \eqref{eq:DK_op} is computationally similar to finding augmented optimal designs. Algorithms that find augmented optimal designs, which are discussed in more detail in Section \ref{subsec:adapt_update}, can be used in most cases to find $\lambda_{j}^{*}(\theta)$. The exception is when $K_{\theta}$ is not positive definite, see the remark at the end of the next subsection for a more detailed discussion. For each run the number of design points, the points themselves and the allocations can change. Practically, depending on the underlying model, finding $\lambda_{j}^{*}(\theta)$ could represent a considerable computational burden. In order to ease this computational burden and eliminate the positive definite requirement the design we develop is an analytic, albeit approximate, solution to $\lambda_{j}^{*}(\theta)$ for runs $j=2,\ldots,J$.


\subsection{The Local Observed Information Adaptive Design (LOAD)}
Define $\xi_{\Psi}^{*}(\theta)$  to be the FLOD with respect to criterion $\Psi$. Denote the design points in $\xi_{\Psi}^{*}(\theta)$ with positive weight as $\bm{x^{*}} = (x_{1}^{*}, \ldots , x_{d}^{*})$ and the corresponding non-zero allocations $\bm{w^{*}} =  (w_{1}^{*}, \ldots , w_{d}^{*})$. Since the approach is local at the same fixed initial guess of $\theta$ for all runs we propose fixing the design points at $\bm{x^{*}}$ and attempt only to alter the allocations to these fixed points.

Some notation regarding sequential experiments. Let $M_{ij} = \sum_{k=1}^{j}m_{ik}$ and $\mathcal{D}_{M_{ij}}$ denote the total number of samples and the data collected at $x_{i}^{*}$ from the first $j$ runs, respectively. We can rewrite equation \eqref{eq:K}, after some basic algebra, as
\begin{align}
K_{\theta}(\lambda_{j},\mathcal{D}_{M_{j-1}}) 
& = \{m_{j} + Q_{\eta}(\mathcal{D}_{M_{j-1}})\} \sum_{i=1}^{d} \zeta_{ij}(w_{ij}) \mu_{\eta}(x_{i}^{*}) f_{x}(x_{i}^{*})f_{x}^{T}(x_{i}^{*}) \\
& = \{m_{j} + Q_{\eta}(\mathcal{D}_{M_{j-1}})\}M_{\theta}\{\nu(\bm{w_{j}})\} \label{eq:Knew},
\end{align}
where $\bm{w_{j}} = (w_{1j},\ldots,w_{dj})$,
\begin{align}
\nu(\bm{w_{j}}) = \left\{\begin{array}{cccc} x_{1}^{*} & \ldots & x_{d}^{*} \\ \zeta_{1j}(w_{1j}) & \ldots & \zeta_{dj}(w_{dj}) \end{array}\right\}
\end{align}
and
\begin{align} \label{eq:rho}
\zeta_{ij}(w_{ij}) = \frac{m_{j}w_{ij}  + Q_{\eta}(\mathcal{D}_{M_{j-1}})\omega_{\eta}(\mathcal{D}_{M_{ij-1}})}{m_{j} + Q_{\eta}(\mathcal{D}_{M_{j-1}})}.
\end{align}
In equation \eqref{eq:Knew}, the term $m_{j} + Q_{\eta}(\mathcal{D}_{M_{j-1}})$ does not depend on the design $\lambda_{j}$. Therefore, equation \eqref{eq:DK_op} can be equivalent written as
\begin{align} \label{eq:DM_opt}
\lambda_{j}^{*}(\theta) = \arg\min_{\lambda_{j} \in \Xi_{\Delta}} \Psi[M_{\theta}\{\nu(\bm{w_{j}})\}],
\end{align}
Finding the solution using \eqref{eq:DM_opt} is likely not a computation simplification to using equation \eqref{eq:DK_op}.

However, on a careful inspection of \eqref{eq:DM_opt} we can further simplify this procedure and obtain an approximate analytic solution for runs $j=2,\ldots,J$. Let $w_{ij}'$ be the solution to $\zeta_{ij}(w_{ij})=w_{i}^{*}$, i.e.
\begin{align} \label{eq:wij}
w_{ij}' = w_{i}^{*} + \frac{Q_{\eta}(\mathcal{D}_{M_{j-1}})\{w_{i}^{*}-\omega_{\eta}(\mathcal{D}_{M_{ij-1}})\}}{m_{j}}.
\end{align}
Let $\lambda_{j}'(\theta)$ be the design with design points $\bm{x^{*}}$ and allocation weights $\bm{w_{j}'} = w_{1j}',\ldots,w_{dj}'$. 
By definition
\begin{align}
M_{\theta}\{\lambda_{j}'(\theta)\} = M_{\theta}\{\xi_{\Psi}^{*}(\theta)\}.
\end{align}
Therefore, if $\lambda_{j}'(\theta)\in\Xi_{\Delta}$ then it is a solution to equation \eqref{eq:DM_opt} and can be used as the design of the $j$th run. Unfortunately, $\lambda_{j}'(\theta)\in\Xi_{\Delta}$ is not always satisfied; since, while it is true that $\sum_{i} w_{ij}' = 1$; it is not guaranteed that $0 \le w_{ij}' \le 1$ for all $i=1,\ldots,d$. We employ a simple fix for this violation by modifying the allocations to
\begin{align}
\tilde{w}_{ij} = \begin{cases} \frac{ w_{ij}'}{ \sum I(w_{ij}'>0)w_{ij}'} & \mbox{ if } w_{ij}'>0 \\ 0 & \mbox{ otherwise } \end{cases}.
\end{align}

Based on this discussion we can now formally state the local design we propose. \\ \mbox{ } \\
\textbf{Local Observed Information Adaptive Design (LOAD)}
\begin{enumerate}
\item{For a fixed initial guess of $\theta$ find $\xi_{\Psi}^{*}(\theta)$ for a desired criterion $\Psi$.}
\item{For run $j=1$ initiate the design by setting $\lambda_{1}(\theta)$, the design of the first stage, to have design points  $x_{1}^{*}, \ldots, x_{d}^{*}$ and corresponding allocations $1/m_{1}$ to each design point. }
\item{For runs $j=2,\ldots,J$ allocate observations to the $i$th optimal design point $x_{i}^{*}$ according to $\tilde{w}_{ij}$.}
\end{enumerate}
Denote the design of the $j$th run in the LOAD as $\tilde{\lambda}_{j}(\theta)$ and the full LOAD as $\tilde{\xi}_{\Psi}(\theta) = \sum_{j} m_{j}\tilde{\lambda}_{j}(\theta)/n$. While $\tilde{\lambda}_{j}(\theta)$ is not necessarily a solution to equation \eqref{eq:DM_opt} we expect, in many cases, it is a small sacrifice in efficiency for a significant reduction in computational complexity.

A final remark on possible computational issues with finding direct solutions to equation \eqref{eq:DK_op}. As stated $J_{\theta}$ is not always positive semi-definite which can result in $K_{\theta}$ being negative definite on the entire design space. In such cases all designs are, by definition, equivalent with respect to $\Psi$. The LOAD indirectly addresses the negative definite cases. Recall the definition of OFI, with support points at the optimal points,
\begin{align}
J_{\theta}(\mathcal{D}_{n}) = \sum_{i=1}^{d} \sum_{j=1}^{n_{i}} I_{\eta}(x^{*}_{i},y_{ij}) f_{x}(x^{*}_{i}) f_{x}^{T}(x^{*}_{i})
\end{align}
and note if $J_{\theta}$ is not positive semi-definite then $\sum_{j=1}^{n_{i}} I_{\eta}(x^{*}_{i},y_{ij}) < 0$ for at least one $i$. The implication is that, in effect, negative information has been collected at these design points. With this in mind we can rewrite \eqref{eq:wij} as
\begin{align}
w_{ij}' = w_{i}^{*}\left\{1 + \frac{Q_{\eta}(\mathcal{D}_{M_{j-1}})}{m_{j}}\right\} - \frac{\sum_{j=1}^{n_{i}} I_{\eta}(x^{*}_{i},y_{ij})}{\mu_{\eta}(x_{i})m_{j}}.
\end{align}
The second term in the right hand side is positive for those $i$ where $\sum_{j=1}^{n_{i}} I_{\eta}(x^{*}_{i},y_{ij}) < 0$. Therefore, the LOAD places additional weight on the design points that currently have negative information thereby increasing the probability that the information at these designs points will be positive after the next run. In this way the LOAD indirectly deals with non positive semi-definite information.

\subsection{Second Order Observed Efficiency}

The main theoretical result regarding the performance of the LOAD is presented in the following theorem and corresponding corollary.
\begin{theorem} \label{thm:Apapt_D_eff}
For any convex optimality criterion and a model satisfying conditions stated in the appendix
\begin{enumerate}[(i)]
\item{if $\mathcal{D}_{n_{i}}^{*}$ is the data following a FLOD then
\begin{align}
\omega_{\eta}(\mathcal{D}_{n_{i}}^{*}) - w_{i}^{*} = O_{p}(n^{-1/2})
\end{align}}
\item{if $\tilde{\mathcal{D}}_{n}$ is the data following the LOAD design and $m_{j}$ is finite, $I_{\eta}:R^{s+1}\rightarrow R^{+}$ and Var$[I_{\eta}(x_{i}^{*},y)]<\infty$, for $i=1,\ldots,d$ and $j=1,\ldots,J$ then
\begin{align}
\omega_{\eta}(\tilde{\mathcal{D}}_{n_{i}}) - w_{i}^{*} = O_{p}(n^{-1})
\end{align}}
\end{enumerate}
for $i=1,\dots,d$ and any $\theta\in\Theta$.
\end{theorem}
Proof is in the Appendix. A corollary regarding the local observed efficiencies is
\begin{corollary} \label{cor:Apapt_D_eff}
For the conditions in Theorem \ref{thm:Apapt_D_eff}
\begin{enumerate}[(i)]
\item{$\Psi_{\rm{eff}}^{\theta}\{\tau_{\eta}(\mathcal{D}_{n}^{*})\} = 1 - O_{p}(n^{-1/2}).$}
\item{$\Psi_{\rm{eff}}^{\theta}\{\tau_{\eta}(\tilde{\mathcal{D}}_{n})\} = 1 - O_{p}(n^{-1}).$}
\end{enumerate}
\end{corollary}
The proof of the corollary is omitted but follows from using Theorem \ref{thm:Apapt_D_eff} in the definitions of local observed $\Psi$-efficiency. Recall, by Lemma \ref{thm:Q} $\Psi_{\rm{eff}}^{\theta}\le 1$. Therefore, according to the corollary the use of a fixed experiment has first order local observed $\Psi$-efficiency; whereas, the LOAD has second order local observed $\Psi$-efficiency.

\section{Estimating the Design After Each Run.} \label{subsec:adapt_update}

The adaptive procedure in the preceding section considered $\theta$ fixed throughout the entire experiment. In cases where EFI depends on $\theta$ a popular alternative to the FLOD is to use an AOD. As previously described an AOD evaluates the FLOD at the MLE obtained from the data of the previous $j-1$ runs, denoted $\hat{\theta}_{M_{j-1}}$. The analytic approach employed in the LOAD is, likely, not appropriate. Instead we must solve equation \eqref{eq:DK_op} evaluated at the MLE directly. In this section we consider both the continuous and exact designs approaches. As stated this is computationally similar to augmented designs which we review in the first part of this section.

\subsection{Design Augmentation}

Suppose an experiment was conducted according to design $\xi_{n_{1}}$ with sample size $n_{1}$. After the data are observed it is clear the \textit{a priori} model was inadequate. For example, higher order terms or interactions were found to be required. In light of the new evidence it is necessary to augment the original design with a new design of size $n_{2}$. For a design $\xi$, the normalized EFI, given the initial design $\xi_{n_{1}}$, is
\begin{align}
M_{\theta}^{\alpha}(\xi) = \alpha M_{\theta}(\xi) + (1-\alpha)M_{\theta}(\xi_{n_{1}}),
\end{align}
where $\alpha = n_{2}/(n_{1} + n_{2})$. Then an optimal augmented design for a general optimality criterion is
\begin{align} \label{eq:Aug_Opt}
\xi_{\Psi}^{*}(\theta) = \arg\min_{\xi \in \Xi} \Psi\{M_{\theta}^{\alpha}(\xi)\},
\end{align}
where $\Xi$ can be either $\Xi_{\Delta}$ or $\Xi_{n_{2}}$, for the continuous and exact approaches, respectively. For continuous $D$-optimal designs the \citet*{Wynn:TheS:1970} algorithm can still be used. Algorithms for exact designs for general criteria are discussed in \citet*{Heib:Dula:Holl:Opti:1993}. Algorithms to find Bayesian designs are computation similar; see \citet*{Chal:Opti:1984}.

\subsection{Adaptive Optimal Design} \label{subsec:AOD}
In adaptive optimal designs (AOD) the \textit{a priori} model is not changed from one run to the next; instead the initial guess at the model parameters is updated based on the MLE from the available data. We begin by defining the basic form of an AOD. Let
\begin{align}
M_{\theta}^{\alpha_{j}}(\lambda_{j}) = \alpha_{j} M_{\theta}(\lambda_{j}) + (1-\alpha_{j})M_{\theta}(\xi_{M_{j-1}}),
\end{align}
where $\alpha_{j}=m_{j}/\{m_{j} + M_{j-1}\}$ and $\xi_{M_{j-1}} = \sum_{k=1}^{j-1} m_{k}\lambda_{k}/M_{j-1}$. An AOD uses
\begin{align}
\lambda_{j}^{*}(\hat{\theta}_{M_{j-1}}) = \arg\max_{\lambda_{j} \in \Xi} \Psi\left\{M_{\hat{\theta}_{M_{j-1}}}^{\alpha_{j}}(\lambda_{j})\right\}
\end{align}
for the design of the $j$th run. This is the basic form of AODs examined in \citet*{Drag:Fedo:Adap:2005}, \citet*{Drag:Husua:Padm:adap:2007}, \citet*{Drag:Fedo:Wu:Adap:2008} and \citet*{Lane:Yao:Flou:Info:2014} and \citet*{Kim:Flou:2014}. We denote the observed design following an AOD procedure as $\overline{\xi}_{\Psi} = \sum_{j} m_{j}\lambda_{j}^{*}(\hat{\theta}_{M_{j-1}})/n$.

\subsection{Estimated Observed Information Design}
The adaptive design proposed in this section is an extension of the LOAD in that it combines the LOAD approach and the AOD approach. Let $\beta_{j} = m_{j}/\{m_{j} + Q_{\eta}(\mathcal{D}_{M_{j-1}})\}$, now equation \eqref{eq:K} can be written, after some basic algebra, as
\begin{align}
K_{\theta}(\lambda_{j},\mathcal{D}_{M_{j-1}}) &= \{m_{j} + Q_{\eta}(\mathcal{D}_{M_{j-1}})\}[\beta_{j} M_{\theta}(\lambda_{j}) + (1-\beta_{j})M_{\theta}\{\tau_{\eta}(\mathcal{D}_{M_{j-1}})\}].
\end{align}
The term $m_{j} + Q_{\eta}(\mathcal{D}_{M_{j-1}})$ is a known constant after run $j-1$. Let
\begin{align}
K_{\theta}^{\beta_{j}}(\lambda_{j},\mathcal{D}_{M_{j-1}}) = \beta_{j} M_{\theta}(\lambda_{j}) + (1-\beta_{j})M_{\theta}\{\tau_{\eta}(\mathcal{D}_{M_{j-1}})\},
\end{align}
so equation \eqref{eq:Aug_Opt} can be written in the context of adaptive OFI designs
\begin{align} \label{eq:Aug_Opt_Adapt}
\lambda_{j}^{*}(\theta) = \arg\min_{\lambda_{j} \in \Xi} \Psi\{K_{\theta}(\lambda_{j},\mathcal{D}_{M_{j-1}})\} = \arg\min_{\lambda_{j} \in \Xi} \Psi\{K_{\theta}^{\beta_{j}}(\lambda_{j},\mathcal{D}_{M_{j-1}})\}.
\end{align}
Now we define a procedure where the solution to \eqref{eq:Aug_Opt_Adapt} evaluated at MLE is found after each successive run. \\ \mbox{ } \\
\textbf{Maximum Likelihood Estimated Observed Information Adaptive Design (MOAD)}
\begin{enumerate}
\item{For a fixed initial guess of $\theta$, find $\xi_{\Psi}^{*}(\theta)$ for an arbitrary criterion $\Psi$.}
\item{For run $j=1$ initiate the design by setting $\lambda_{1}(\theta)$, the design of the first stage, to have design points  $x_{1}^{*}, \ldots, x_{d}^{*}$ and corresponding allocations $1/m_{1}$ to each design point.}
\item{For run $j=2,\ldots,J$ compute $\hat{\theta}_{M_{j-1}}$ and allocate observations according to the design
\begin{align}
\lambda_{j}^{*}(\hat{\theta}_{M_{j-1}}) = \arg\min_{\lambda_{j} \in \Xi} \Psi\left\{K_{\hat{\theta}_{M_{j-1}}}^{\beta_{j}}(\lambda_{j},\mathcal{D}_{M_{j-1}})\right\},
\end{align}
where $\Xi=\Xi_{\Delta}$ or $\Xi_{m_{j}}$ for the continuous and exact designs approach, respectively.}
\end{enumerate}

\citet*{Barn:Sore:ARev:1994} point out that the OFI evaluated at the MLE, $J_{\hat{\theta}_{n}}$, is always positive semi-definite. Therefore, there is no non-positive definite issues with the MOAD, provided the MLE exists. As a result, existing procedures to find augmented designs can be directly used to solve \eqref{eq:Aug_Opt_Adapt}. Denote the observed MOAD as $\hat{\xi}_{\Psi} = \sum_{j} m_{j}\lambda_{j}^{*}(\hat{\theta}_{M_{j-1}})/n$.

\subsection{Variance Reduction versus Reduced Variance Approximation} \label{subsec:VarApprox}

We close this section by comparing the two proposed design procedures. From a design perspective the use of an initial guess of $\theta$ is a classical approach. However, in analysis it is not appropriate to approximate the OFI at an initial guess. Instead, we use the inverse of OFI evaluated at the MLE, $J_{\hat{\theta}_{n}}^{-1}$.

Considering the analysis perspective leads to nuance in the selection of LOAD or MOAD. The OFI measures the information locally in the neighborhood of $\theta$. As such the accuracy and efficiency of the MLE is most closely related to $J_{\theta}^{-1}$. As a result the LOAD will, likely, lead to more precise estimates. On the other hand, we approximate the variance of the MLE with $J_{\hat{\theta}_{n}}^{-1}$. The MOAD will, likely, lead to a reduction in the variance approximation. The question is then, do we want to reduce the variance of the MLE or its approximation?  We will examine this contrast further in the examples of Section \ref{sec:GHR}.



\section{Examples} \label{sec:GHR}

In this section we consider two simulations studies from the class of generalized linear models. Recall the shorthand and notation used are collected in Table \ref{tab:notatoin}.

\subsection{Gamma Regression with Log Link}
Consider a gamma regression model with a log link which has responses defined by density
\begin{align}
f(y|\eta) = \frac{1}{y\Gamma(\alpha)}(y\alpha e^{-\eta})^{\alpha}e^{-y\alpha e^{-\eta}}, \mbox{ } y \ge 0,
\end{align}
where $\eta = \theta^{T} f_{x}(x)$ and we assume $\alpha$ is a known constant. It is straightforward to derive OFI and EFI with respect to $\theta$ as
\begin{align}
J_{\theta}(\mathscr{D}_{n}) &= \alpha\sum_{i=1}^{d} n_{i} \overline{y}_{i} e^{-\theta^{T} f_{x}(x_{i})} f_{x}(x_{i})f_{x}^{T}(x_{i}) \\
\end{align}
and
\begin{align}
M(\xi_{n}) = \alpha\sum_{i=1}^{d} w_{i}f_{x}(x_{i})f_{x}^{T}(x_{i}).
\end{align}
This example is intriguing since EFI does not depend on $\theta$; but, the OFI does. Since EFI does not depend on $\theta$, the FLOD is considered global, over the entire parameter space, and AODs are not applicable. We still refer to the global optimal design as the FLOD to be consistent with the remainder of the text. Note that EFI is proportional, up to the constant $\alpha$, to the EFI of a traditional linear model.

For the simulation study we let $f_{x}(x) = (1, x_{1} , x_{2})$, where $x_{1}$ and $x_{2}$ are two independent factors; $\theta = (1,1,1)$; and the design space vertices of $[-1,1]^{2}$, i.e. $\mathcal{X} = \{ (1,1), (1,-1), (-1,1),(-1,-1)\}$. For this model the continuous $D$ and $A$-optimal designs are the same and place equal weight on the $d=4$ vertices.

In this simulation study we consider various sample sizes, $n$, that are multiples of $d$; to ensure the exact and continuous FLOD are equivalent. The LOAD and MOAD were initialized with a run of size 4, i.e., $m_{1}=4$, with 1 observation at each of the 4 optimal design points. Each subsequent run size is 1, i.e., $m_{j}=1$, $j=2,\ldots,J$. Due to the simplicity of the underlying model, we used the exact version of the MOAD in all simulations. 

There are three aspects of the adaptive designs that will be evaluated; the local observed efficiency, $\Psi_{\rm{eff}}^{\theta}$; the approximate observed efficiency, $\Psi_{\rm{eff}}^{\hat{\theta}_{n}}$; and the relative efficiency of the unconditional variance, $\mbox{Rel-Eff}_{\Psi}$. Neither the LOAD or the MOAD directly target the unconditional variance; however, it is possible that if the conditional variance of the MLE is improved then the unconditional variance will also improved.

\subsubsection{Local Observed Efficiency}

The OFI evaluated locally at the true parameter represents the benchmark measure of information [\citet*{Barn:Sore:ARev:1994}]. The primary objective of this work has been to develop adaptive designs that maximize this local efficiency. This measure of efficiency depends on the observed data. To compare designs, we must compare not only median efficiency, but also its variability.

The top row of Figure \ref{fig:Gamma} presents box and whisker plots (of the 5th, 10th, 25th, 50th, 75th and 95th percentiles), obtained from 10,000 simulations,  of $\Psi_{\rm{eff}}^{\theta}$ for the $D$ and $A$ criteria. Each column represents a different sample size from left to right $n=12,36,100$. In each figure results are given for the FLOD (white), the LOAD (light gray) and the MOAD (gray). The shape parameter $\alpha=0.1$ for all simulations. The median $\Psi_{\rm{eff}}^{\theta}$ of the LOAD is greater and its variability is reduced compared to both the FLOD and the MOAD. This is true for each sample size and design criterion as predicted by Theorem \ref{thm:Apapt_D_eff}. Further, the MOAD outperforms the FLOD in all cases considered. For example, when $n=36$ the median [interquartile range (IQR)] with resect to $A_{\rm{eff}}^{\theta}$ for the LOAD [0.90 (0.79-0.96)] is better than the MOAD [0.72 (0.54-0.86)] and the FLOD [0.63 (0.46-0.77)]. The superior performance of LOAD is, at least in part, due to its execution at the true parameter $\theta$. The MOAD does not use the true parameter. These simulations suggest a similar conclusion regarding the LOAD versus the MOAD as \cite{Dett:Born:OnTh:2013} and \citet*{Lane:Yao:Flou:Info:2014} found when examining the AOD and the FLOD. That is, a local guess often performs better than updating for small and moderate sample sizes; and the better the guess the larger the sample size required for the updating procedure to perform better. The difference here is that in terms of $\Psi_{\rm{eff}}^{\theta}$ the FLOD is never the best performer of the three designs considered [the FLOD, the MOAD and the  LOAD] and the AOD is not applicable. A general guideline would be to use the LOAD when a the a priori information regarding the true value of $\theta$ is strong and to use the MOAD otherwise.  We conducted the same simulations for varying values of $\alpha$. These are not included since, in general, increasing $\alpha$ yielded similar results to increasing the sample size.

\subsubsection{Observed Efficiency at the MLE}

The efficiency of the OFI at the MLE, $\Psi_{\rm{eff}}^{\hat{\theta}_{n}}$, is also an important benchmark for the performance a design. After an experiment is completed, confidence intervals and hypothesis tests will be generated using the entries of the $J_{\hat{\theta}_{n}}^{-1}$. Increasing  $\Psi_{\rm{eff}}^{\hat{\theta}_{n}}$ will lead to narrower confidence intervals and more powerful tests.

\begin{figure}
\centering
\begin{subfigure}[h]{0.32\textwidth}
    \centering
    \includegraphics[scale=.58]{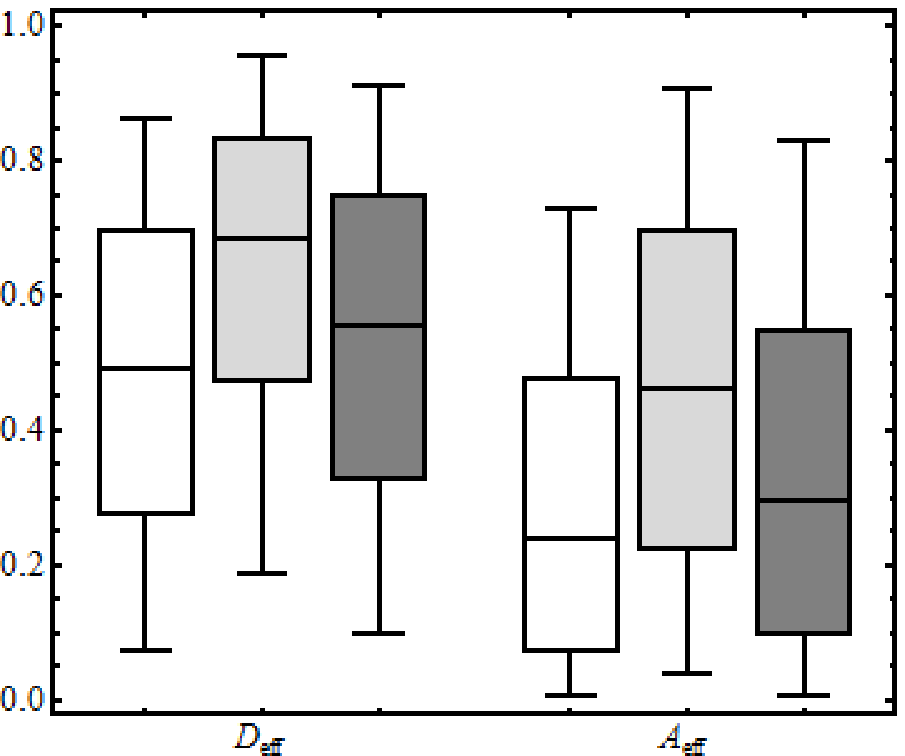}
\end{subfigure} %
\begin{subfigure}[h]{0.32\textwidth}
    \centering
    \includegraphics[scale=.58]{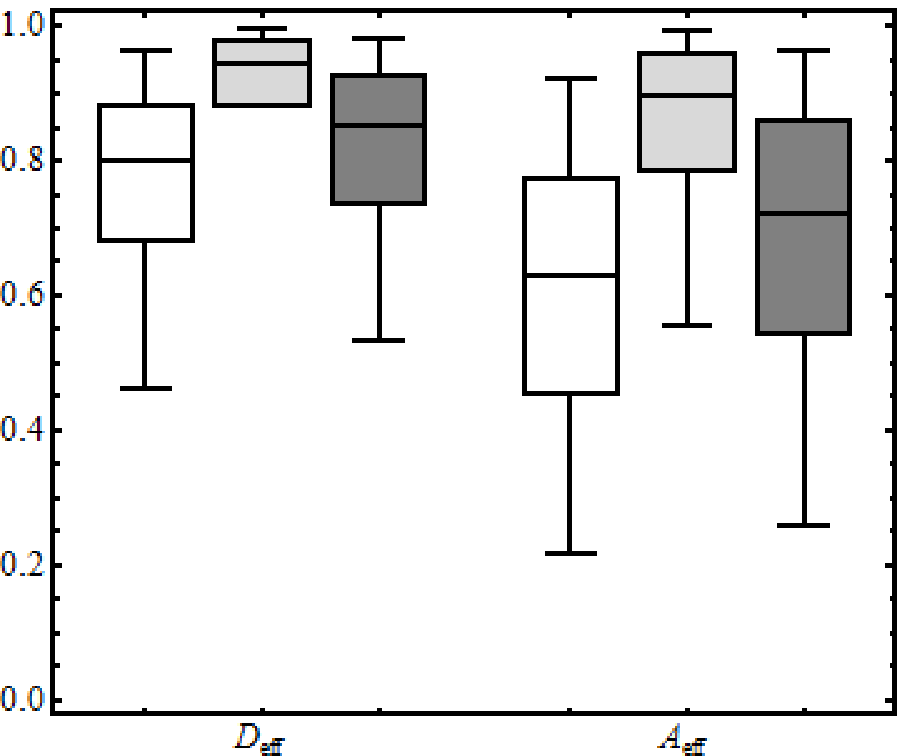}
\end{subfigure} %
\begin{subfigure}[h]{0.32\textwidth}
    \centering
    \includegraphics[scale=.58]{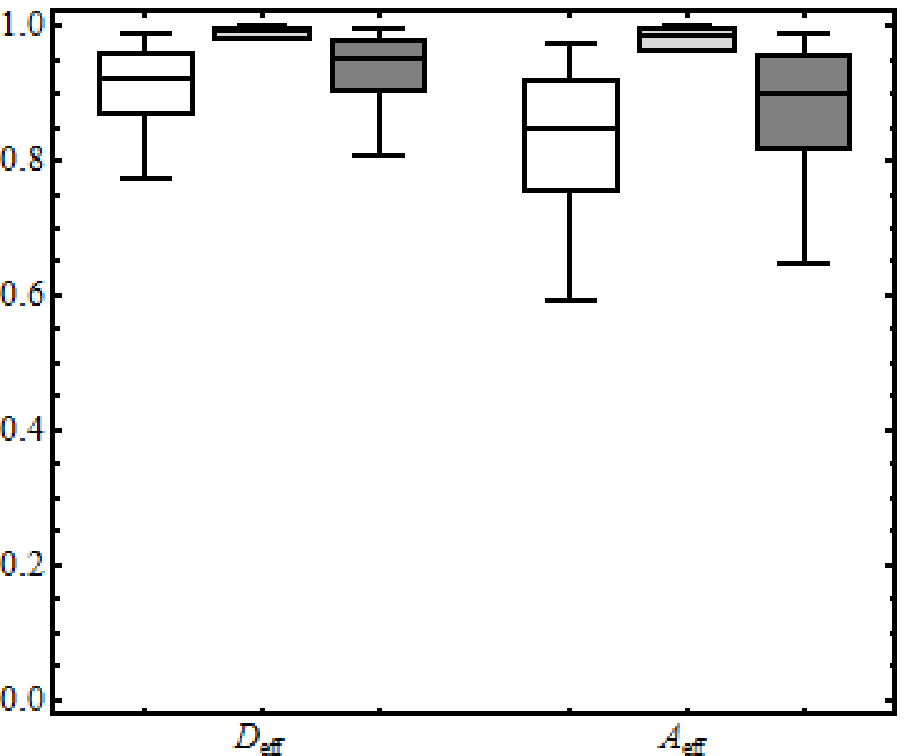}
\end{subfigure} \\
\begin{subfigure}[h]{0.32\textwidth}
    \centering
    \includegraphics[scale=.58]{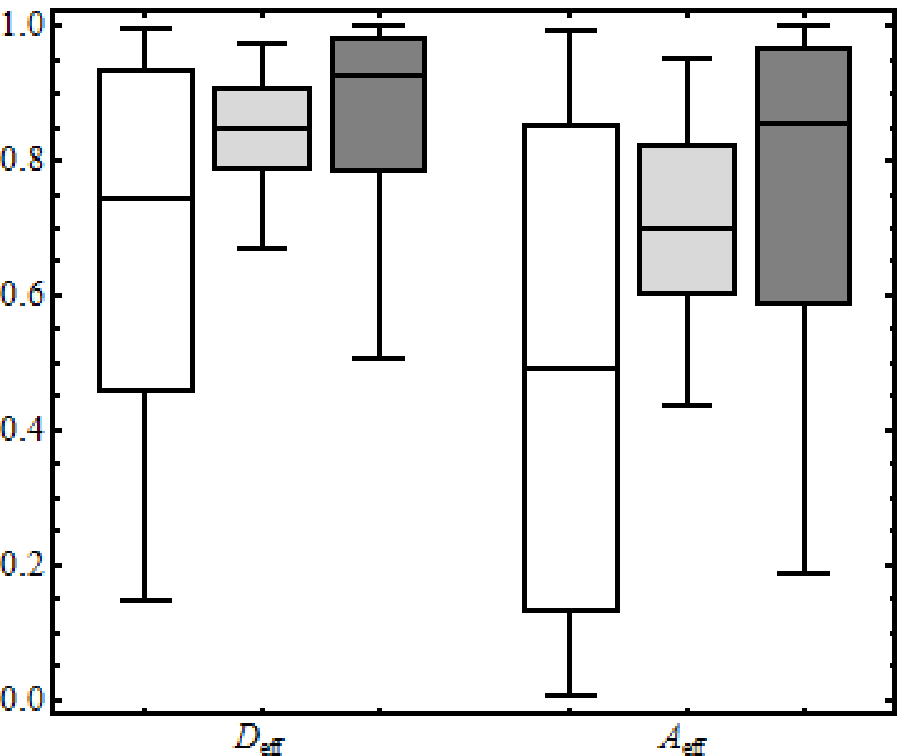}
\end{subfigure} %
\begin{subfigure}[h]{0.32\textwidth}
    \centering
    \includegraphics[scale=.58]{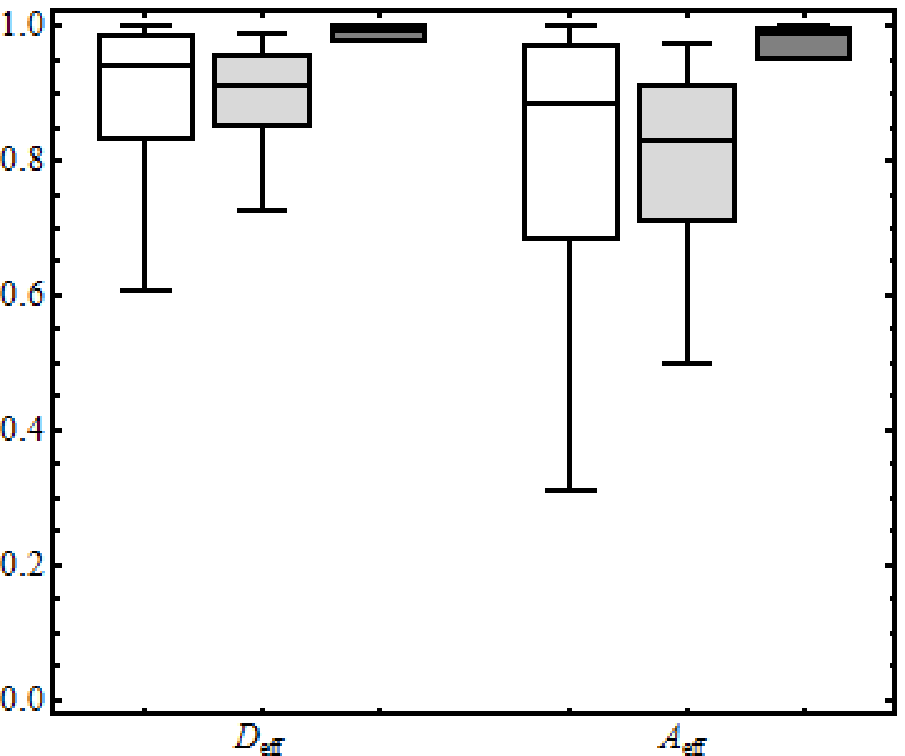}
\end{subfigure} %
\begin{subfigure}[h]{0.32\textwidth}
    \centering
    \includegraphics[scale=.58]{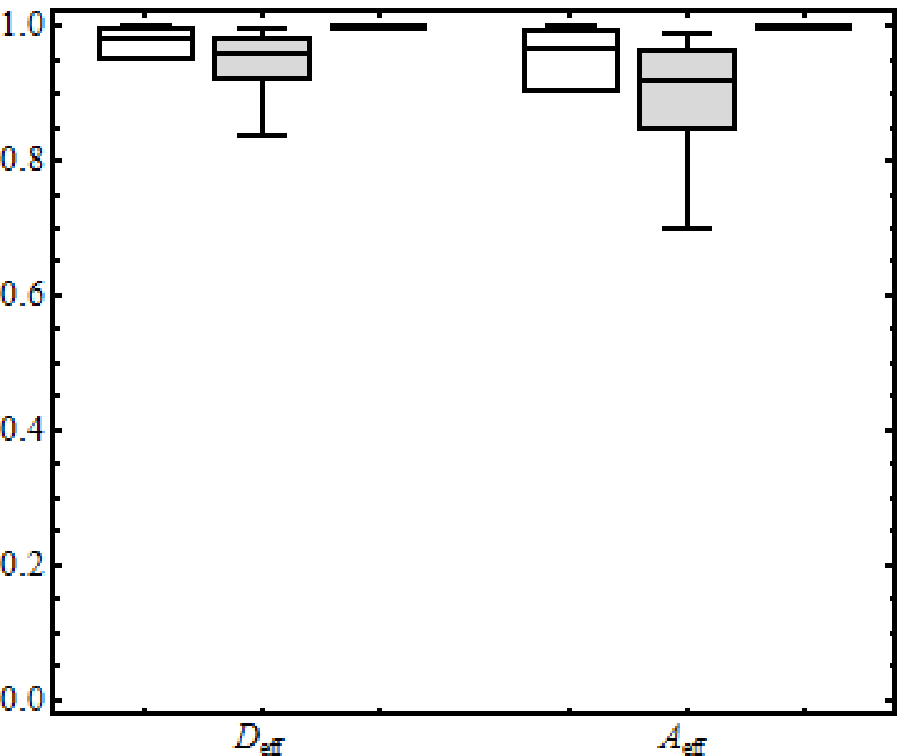}
\end{subfigure}
\caption{Box and whisker plots (of the 5th, 10th, 25th, 50th, 75th and 95th percentiles), from 10,000 simulations, of the $\Psi_{\rm{eff}}^{\theta}$ (top) and $\Psi_{\rm{eff}}^{\hat{\theta}_{n}}$ (bottom) for the FLOD (white), the LOAD (light gray) and the MOAD (gray). Each column represents a different sample size from left to right $n=12,36,100$.}\label{fig:Gamma}
\end{figure}

Box and whisker plots of $\Psi_{\rm{eff}}^{\hat{\theta}_{n}}$ are presented in the bottom row of Figure \ref{fig:Gamma} for the same simulations as previously discussed. The MOAD results in a greater median and reduced variability in $\Psi_{\rm{eff}}^{\hat{\theta}_{n}}$ compared to all other designs. For example, when $n=36$ the median (IQR) of $A_{\rm{eff}}^{\hat{\theta}_{n}}$ for the MOAD [0.99 (0.95-1.00)] is greater than the FLOD [0.88 (0.68-0.97)] and the LOAD [0.83 (0.71-0.91)]. The LOAD performs slightly better both in terms of the median and IQR than the FLOD when $n=12$. However, as the sample size increases the FLOD has a slightly greater median; however, the IQR remains comparable.

The results for the local and observed efficiency support the discussion in Section \ref{subsec:VarApprox}. The MOAD preformed the best in terms of increasing the $\Psi_{\rm{eff}}^{\hat{\theta}_{n}}$; whereas the LOAD performed the best in terms of $\Psi_{\rm{eff}}^{\theta}$ (see Figure \ref{fig:Gamma}).

\subsubsection{Unconditional Efficiency of the Variance}

This preceding discussion provides evidence that distribution of the Var$[\hat{\theta}|a]$ is improved using the two proposed adaptive procedures. As stated in the introduction, the present author shares the prevailing opinion that, in general, conditional inference is preferable to unconditional inference. Therefore, the preceding comparisons of $\Psi_{\rm{eff}}^{\theta}$ and $\Psi_{\rm{eff}}^{\hat{\theta}_{n}}$ provide sufficient evidence in support of the LOAD or MOAD in place of the FLOD.  However, intuitively, these conditional improvements might translate to an improvement in the unconditional variance of the MLE.

To assess the unconditional efficiency of the MLE of each adaptive procedure we compare its efficiency relative to the MLE of the FLOD. Here we define the unconditional $D$ and $A$ efficiency of the LOAD and MOAD designs, represented by ADAPT below, relative to the FLOD as
\begin{align}
\mbox{Rel-Eff}_{D} = \frac{|\mbox{Var}[\hat{\theta}(\xi_{\mbox{flod}})]|^{1/p}}{|\mbox{Var}[\hat{\theta}(\xi_{\mbox{adapt}})]|^{1/p}} \quad \mbox{ and } \quad  \mbox{Rel-Eff}_{A} = \frac{\mbox{Tr}\{\mbox{Var}[\hat{\theta}(\xi_{\mbox{flod}})]\}}{\mbox{Tr}\{\mbox{Var}[\hat{\theta}(\xi_{\mbox{adapt}})]\}},
\end{align}
respectively. As defined, an efficiency greater than 1 indicates that an adaptive design is more efficient than the FLOD. Table \ref{tab:Gamma} presents the relative
efficiency (obtained via simulation) for each of the cases considered in Figure \ref{fig:Gamma}. In every simulation considered the LOAD has the greater efficiency than the FLOD and MOAD. In many cases the improvement in efficiency is significant, with nearly a 70\% increase in relative efficiency in the most extreme examples. Further, the MOAD was more efficient than FLOD in every simulation. 

\begin{table}
\centering
\scriptsize
\begin{tabular}{ ccccc }
& \multicolumn{2}{c}{$\mbox{Rel-Eff}_{D}$} & \multicolumn{2}{c}{$\mbox{Rel-Eff}_{A}$}  \\
\noalign{\smallskip}\hline\hline\noalign{\smallskip}
$n$ & LOAD & MOAD & LOAD & MOAD \\
\noalign{\smallskip}\hline\noalign{\smallskip}
12 & 1.68 & 1.24 & 1.69 & 1.24 \\
36 & 1.32 & 1.14 & 1.33 & 1.15 \\
100 & 1.05 & 1.06 & 1.05 & 1.06 \\
\noalign{\smallskip}\hline\noalign{\smallskip}
\end{tabular}
\caption{Results from 10,000 simulations of the efficiency of the LOAD and the MOAD relative to the FLOD.} \label{tab:Gamma}
\end{table}

\subsection{Normal Regression with a Power Link}

In this section, we consider a second example from the class of generalized linear models. This time a model with normal errors and square root link, i.e.,
\begin{align}
y = \eta^{2} + \varepsilon,
\end{align}
where again $\eta = \theta^{T} f_{x}(x)$, $\varepsilon \sim N(0,\sigma^{2})$ and $\sigma$ is a known constant. It is straightforward to derive OFI and EFI with respect to $\theta$ as
\begin{align}
J_{\theta}(\mathscr{D}_{n}) &= \frac{2}{\sigma^{2}} \sum_{i=1}^{d} n_{i} \left[ 3\{\theta^{T} f_{x}(x_{i})\}^{2} - \overline{y}_{i} \right]  f_{x}(x_{i})f_{x}^{T}(x_{i}) \\
\end{align}
and
\begin{align}
M_{\theta}(\xi_{n}) = \frac{4}{\sigma^{2}} \sum_{i=1}^{d} w_{i} \{\theta^{T} f_{x}(x_{i})\}^{2}  f_{x}(x_{i})f_{x}^{T}(x_{i}).
\end{align}
In this example, both measures of information depend on the model parameters. We examine the same designs as in the preceding example, as well as the AOD procedure described in Section \ref{subsec:AOD}.

As before we use $f_{x}(x) = (1, x_{1} , x_{2})$, where $x_{1}$ and $x_{2}$ are two independent factors. For this simulation we let $\theta = (1,1,1)$ and $\sigma=5$. The design space, $\mathcal{X}$, used in this section is again defined as the vertices of the square $[-1,1]^{2}$. For each sample size the exact $D$ and $A$ FLOD (locally at the true parameters) were found by searching over $\mathcal{X}$ using the \pkg{OptimalDesign} package in R [\citet*{Filo:Harm:Opti:2016}]. Both the $D$ and $A$ optimal designs have positive support on all four points of the design space. Similarly, the true value of the parameters were used for the LOAD. The values of the true parameters are unknown in practice and the FLOD and LOAD results are, perhaps, upper limits on performance. The MOAD and the AOD do not use knowledge of the true parameters and are thus practical representations of performance.

The remainder of this section we discuss the results of the simulation in detail for the same three measures of efficiency described in for the gamma regression simulation.

\subsubsection{Local Observed Information at the True Parameter}

\begin{figure}
\centering
\begin{subfigure}[h]{0.32\textwidth}
    \centering
    \includegraphics[scale=.58]{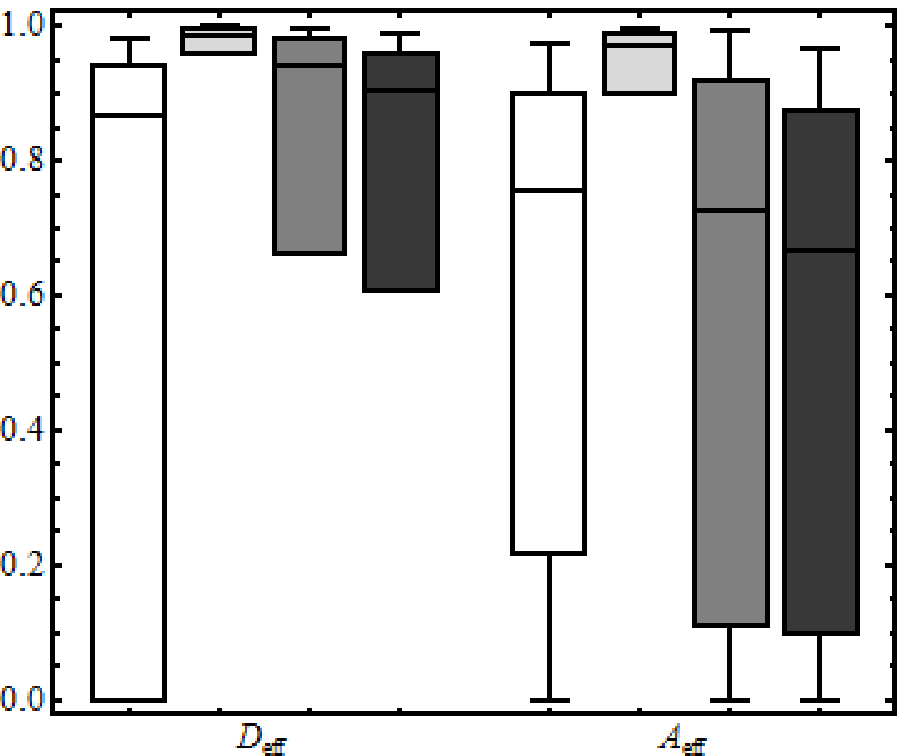}
\end{subfigure} %
\begin{subfigure}[h]{0.32\textwidth}
    \centering
    \includegraphics[scale=.58]{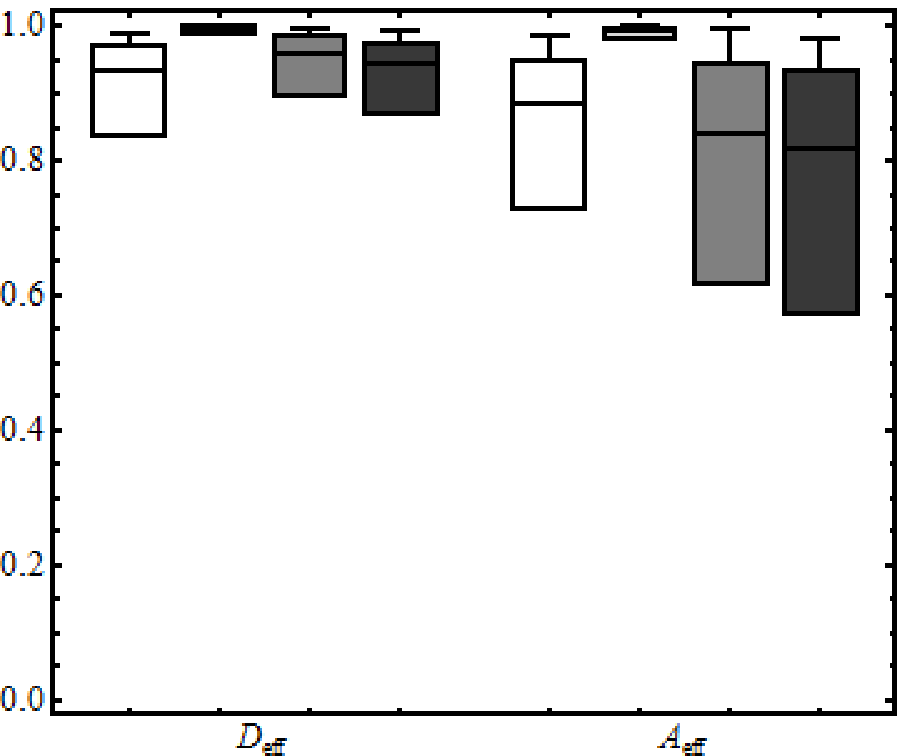}
\end{subfigure} %
\begin{subfigure}[h]{0.32\textwidth}
    \centering
    \includegraphics[scale=.58]{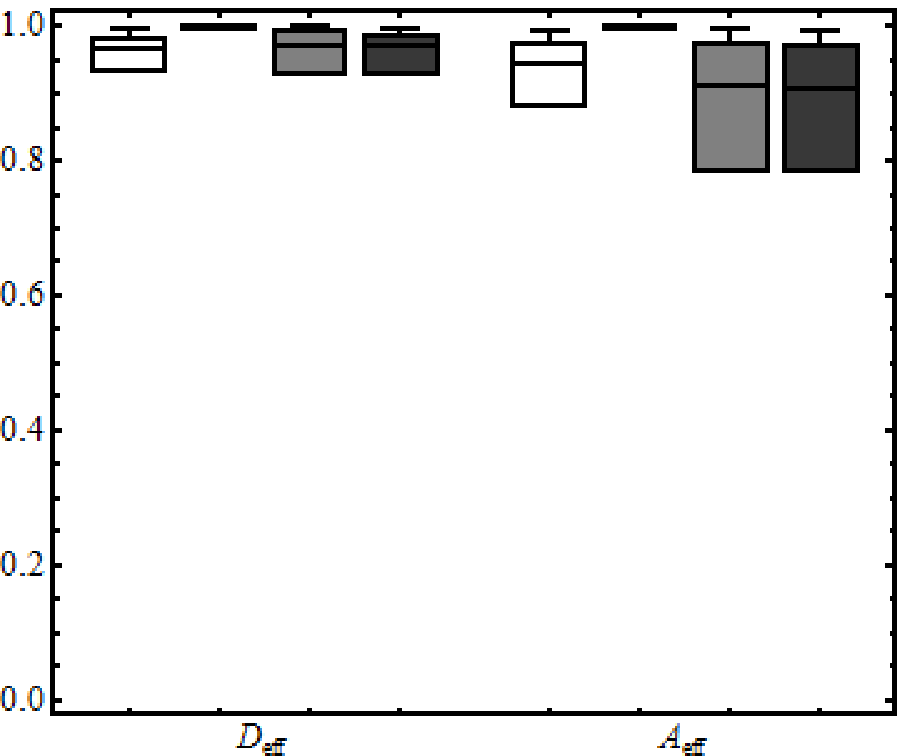}
\end{subfigure} \\
\begin{subfigure}[h]{0.32\textwidth}
    \centering
    \includegraphics[scale=.58]{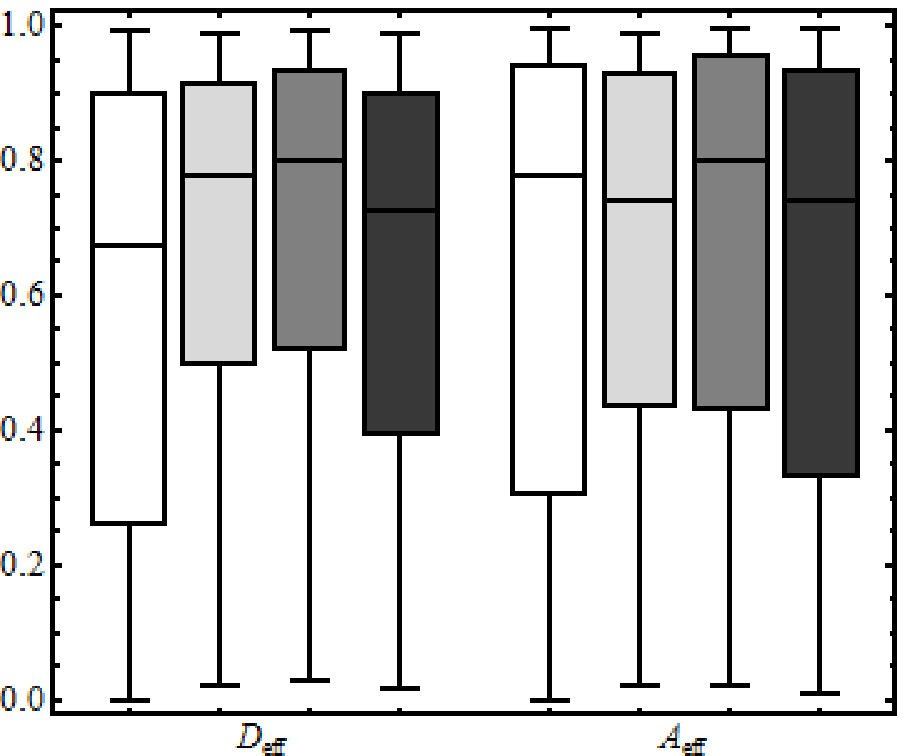}
\end{subfigure} %
\begin{subfigure}[h]{0.32\textwidth}
    \centering
    \includegraphics[scale=.58]{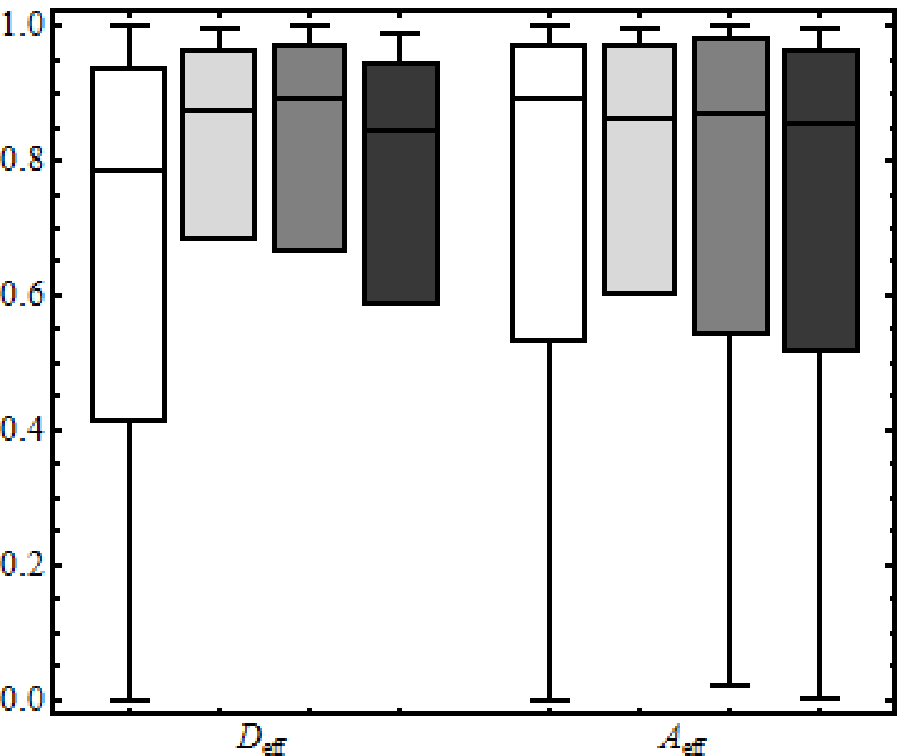}
\end{subfigure} %
\begin{subfigure}[h]{0.32\textwidth}
    \centering
    \includegraphics[scale=.58]{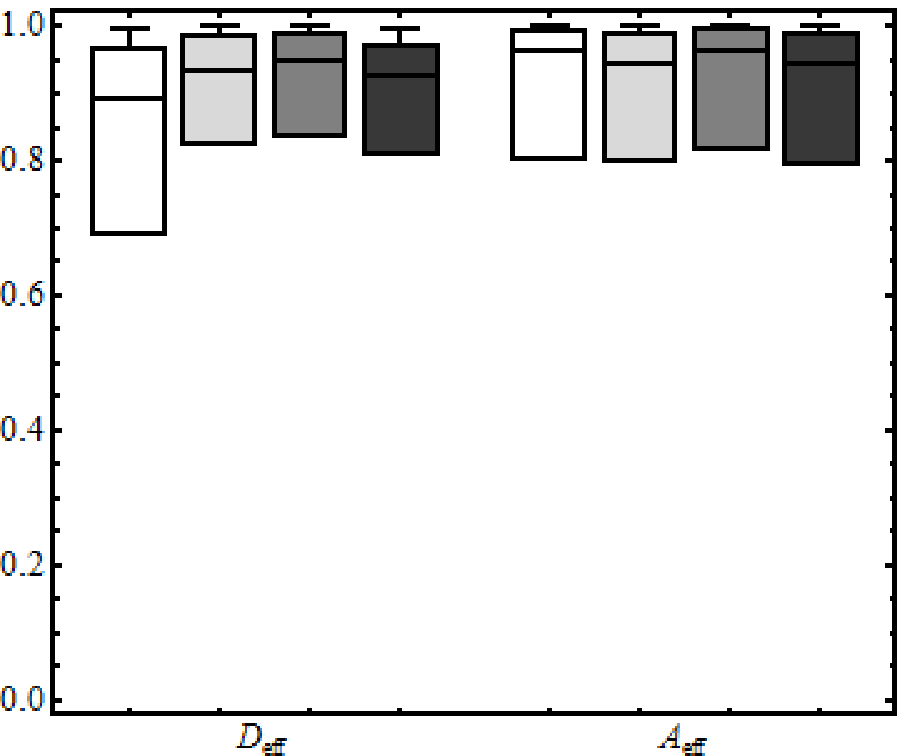}
\end{subfigure}
\caption{Box and whisker plots (of the 5th, 10th, 25th, 50th, 75th and 95th percentiles), from 10,000 simulations, of the $\Psi_{\rm{eff}}^{\theta}$ (top) and $\Psi_{\rm{eff}}^{\hat{\theta}_{n}}$ (bottom) for the FLOD (white), the LOAD (light gray) and the MOAD (gray). Each column represents a different sample size from left to right $n=12,36,100$.}\label{fig:Normal}
\end{figure}

The top row of Figure \ref{fig:Normal} presents box and whisker plots of the $\Psi_{\rm{eff}}^{\theta}$ for the $D$ and $A$ criteria. Each column represents a different sample size from left to right $n=25,50,100$. In each figure results are given for the FLOD (white), the LOAD (light gray), the MOAD (gray) and the AOD (dark gray). The results of this simulation are, in general, comparable to the gamma regression simulation. Once again the median $\Psi_{\rm{eff}}^{\theta}$ of the LOAD is greater and its variability is significantly reduced compared to all three alternative designs.  This is true for each sample size and design criterion. 

\subsubsection{Observed Information at the MLE}

Box and whisker plots of $\Psi_{\rm{eff}}^{\hat{\theta}_{n}}$ are presented in the bottom row of Figure \ref{fig:Normal} for the same simulations previously discussed. The MOAD results in a greater median and reduced variability in $\Psi_{\rm{eff}}^{\hat{\theta}_{n}}$ compared to the FLOD, the LOAD and the AOD which had similar performance to each other with respect to this measure of efficiency.



\subsubsection{Unconditional Efficiency of the Variance}

Recall that a $\mbox{Rel-Eff}_{\Psi}$ greater than one indicates that the MLE following an adaptive designs is more efficient than the MLE following the FLOD in terms of the unconditional variance. Table \ref{tab:Normal} presents the relative efficiency (obtained via simulation) for each of the cases considered in Figure \ref{fig:Normal}, where (1) is the LOAD, (2) is the MOAD and (3) is the AOD. In every simulation considered the LOAD has the greatest relative efficiency. The MOAD was more efficient than both FLOD and the AOD in nearly every simulation. For $A$ optimality the results are in line with expectations; the AOD is significantly less efficient than the FLOD, particularly when the sample size is small. Unexpectedly, the AOD was more slightly more efficient than the FLOD for the $D$ criteria.


\begin{table}
\centering
\scriptsize
\begin{tabular}{ ccccccc }
& \multicolumn{3}{c}{$\mbox{Rel-Eff}_{D}$} & \multicolumn{3}{c}{$\mbox{Rel-Eff}_{A}$}  \\
\noalign{\smallskip}\hline\hline\noalign{\smallskip}
$n$ & LOAD & MOAD & AOD & LOAD & MOAD & AOD \\
\noalign{\smallskip}\hline\noalign{\smallskip}
\multicolumn{7}{c}{Variance} \\
\noalign{\smallskip}\hline\noalign{\smallskip}
25 & 1.47 & 1.35 & 1.24 & 1.45 & 1.14 & 1.00 \\
50 & 1.55 & 1.40 & 1.26 & 1.47 & 1.17 & 1.03 \\
100 & 1.44 & 1.32 & 1.18 & 1.32 & 1.08 & 0.99 \\
\noalign{\smallskip}\hline\noalign{\smallskip}
\end{tabular}
\caption{Results from 10,000 simulations of the efficiency of the LOAD, the MOAD and the AOD relative to the FLOD.} \label{tab:Normal}
\end{table}

\section{Discussion}

This paper demonstrates that incorporating OFI into adaptive designs can improve the efficiency of experiments. Two adaptive procedures were developed, one local and one using the MLE based on the previously collected data. These procedures were contrasted against classical fixed locally optimal designs (FLODs) and adaptive optimal designs (AODs). The local procedure, LOAD, was shown to have second order local efficiency, whereas the FLOD is only first order locally efficient. A simulation study was conducted using two popular generalized linear models. The results of the simulation study provided further evidence that the proposed adaptive procedures lead to increased efficiency against relevant alternatives. Of course, this simulation study is by no means exhaustive and more investigation is needed to understand the full implications of such procedures.

Inference following adaptive designs has received significant attention [see  \citet{Lane:Flou:2012}, \citet{Lane:Yao:Flou:Info:2014}, and \citet{Lane:Wang:Flou:Cond:2016}]. Similar investigation is warranted for the procedures proposed in this work. However; these same concerns might not always be a significant issue for the LOAD and MOAD. For example, models in which the OFI is an ancillary random matrix are not likely to result in poorer inferential properties.

We considered a linear parameterization primarily for clarity of exposition. Nonlinear extensions are expected to be straightforward in principle, although there may exist some unique computation or technical complications. In Section \ref{sec:elem} we related the difference between OFI and EFI to the elemental information. We considered a univariate response. From \citet{Atki:Fedo:Herz:Elem:2014} it can be seen that OFI and EFI will be a function of elemental information even for a multivariate response. This suggests that it is feasible to develop similar adaptive procedures to those proposed in this work for multivariate responses.

We considered only frequentist models and designs. However, it is known that the inverse of OFI is often a better approximation to the variance of the posterior distribution [\citet{Berg:Stat:1985}]. For this reason, adopting the proposed procedures in the Bayesian framework is of interest, but has not been considered here.

\section*{Acknowledgement}

The author would like to thank Radoslav Harman for his helpful comments and suggestions.

\appendix

\section{Proof of Theorem \ref{thm:Apapt_D_eff}}

We begin with the proof of part $i$. For full conditions see Section 8 \citet{Efro:Hink:Asse:1978}. Lemma 2 in \citet{Efro:Hink:Asse:1978} directly implies
\begin{align}
\sqrt{n_{i}} \left\{ q_{\eta}(\mathscr{D}_{n_{i}})/n_{i} - 1\right\} \sim N\left\{0,\sigma_{\eta}^{2}(x_{i})\right\},
\end{align}
where $\sigma_{\eta}^{2}(x) = \mbox{Var}[q_{\eta}(x,y)]$. Due to the independence of responses it can be shown that $\sqrt{n}\{q_{\eta}(\mathcal{D}_{n_{1}})/n-w_{1},\ldots,q_{\eta}(\mathcal{D}_{n_{d}})/n-w_{d}\}$ is asymptotically jointly multivariate normal with mean 0 and a diagonal variance-covariance matrix with $i$th element of the diagonal equal to $w_{i}\sigma_{\eta}^{2}(x_{i})$. Recall $\omega_{\eta}(\mathcal{D}_{n_{i}}) = q_{\eta}(\mathcal{D}_{n_{i}})/\sum_{i} q_{\eta}(\mathcal{D}_{n_{i}})$ therefore by an application of the delta method
\begin{align}
\sqrt{n}\{\omega_{\eta}(\mathcal{D}_{n_{i}}) - w_{i}\} \sim N\left\{0,w_{i}(1-2w_{i})\sigma_{\eta}^{2}(x_{i}) + w_{i}^{2}\sum_{k=1}^{d} w_{k}\sigma_{\eta}^{2}(x_{k}) \right\}
\end{align}
as $n\rightarrow\infty$. The above implies part $i$ of the theorem, i.e.,
\begin{align}
\omega_{\eta}(\mathcal{D}_{n_{i}}) - w_{i} = O_{p}(n^{-1/2}).
\end{align}

The proof of part $ii$ of the Theorem will be a direct result of the following Lemma.
\begin{lemma}
Let $\overline{\mathcal{D}}_{n}$ be the data following any adaptive design such that
\begin{align}
m_{ij} = \begin{cases} a_{ij} & \mbox{ if } w'_{ij}>0 \\ 0 & \mbox{ otherwise }  \end{cases}, \quad i=1,\ldots,d
\end{align}
where $a_{ij}\in \{0,\ldots,m_{j}\}$ is the integer sample size at $x_{i}^{*}$ for run $j$ and $m_{j} = \sum_{i} a_{ij}$. Further assume $\overline{\mathcal{D}}_{n}$ was obtained for any convex optimality criterion from the stated model such that the standard regularity conditions and additional conditions that $m_{j}$ is finite,  $I_{\eta}(x_{i}^{*},y)>0$ and Var$[I_{\eta}(x_{i}^{*},y)]<\infty$ for $i=1,\ldots,d$ are satisfied. The following holds
\begin{align}
\omega_{\eta}(\overline{\mathcal{D}}_{n_{i}}) - w_{i}^{*} = O_{p}(n^{-1}).
\end{align}
for all $i=1,\ldots,d$ and any $\theta\in\Theta$.
\end{lemma}

Proof of the Lemma. Recall the definition of $w'_{i,J+1}$ in equation \eqref{eq:wij}
\begin{align}
w'_{i,J+1} = w_{i}^{*} + Q_{\eta}(\overline{\mathcal{D}}_{M_{J}})\left\{w_{i}^{*} -  \omega_{\eta}(\overline{\mathcal{D}}_{i,M_{J}})\right\}.
\end{align}
First, we consider the set $i_{-} = \{i=1,\ldots,d:w'_{i,J+1}\le0\}$. Denote the last run such that $w'_{ij}>0$ as
\begin{align}
b_{i} =\arg\max_{j} \{j:w'_{ij}>0\}
\end{align}
for $i\in i_{-}$. From the definition of $w'_{ij}$ in Equation \eqref{eq:wij}; since $w_{i,b_{i}}'>0$
\begin{align} \label{eq:Cond1}
w_{i}^{*} - \omega_{\eta}(\overline{\mathcal{D}}_{M_{i,b_{i}-1}}) > -w_{i}^{*}\frac{m_{b_{i}}}{Q_{\eta}(\overline{\mathcal{D}}_{M_{b_{i}-1}})};
\end{align}
and since $w'_{i,J+1}\le 0$
\begin{align} \label{eq:Cond2}
w_{i}^{*} - \omega_{\eta}(\overline{\mathcal{D}}_{M_{i,J}}) \le -w_{i}^{*}\frac{1}{Q_{\eta}(\overline{\mathcal{D}}_{M_{J}})} \le 0
\end{align}
for $i\in i_{-}$. Further, since $w'_{ij}\le0$ for all $j>b_{i}+1$ there are no observations allocated to any $i\in i_{-}$ for $j>b_{i}+1$ which implies we can write
\begin{align}
\omega_{\eta}(\overline{\mathcal{D}}_{M_{i,J}}) &= \frac{q_{\eta}(\overline{\mathcal{D}}_{M_{i,b_{i}-1}}) - q_{\eta}(\overline{\mathcal{D}}_{m_{i,b_{i}}})}{Q_{\eta}(\overline{\mathcal{D}}_{M_{J}})} \\
&= \frac{\omega_{\eta}(\overline{\mathcal{D}}_{M_{i,b_{i}-1}})}{1 + \sum_{j=b_{i}}^{J} Q_{\eta}(\overline{\mathcal{D}}_{m_{j}}) / Q_{\eta}(\overline{\mathcal{D}}_{M_{b_{i}-1}})} - \frac{q_{\eta}(\overline{\mathcal{D}}_{m_{i,b_{i}}})}{Q_{\eta}(\overline{\mathcal{D}}_{M_{J}})} \\
&\le \omega_{\eta}(\overline{\mathcal{D}}_{M_{i,b_{i}-1}}) \\
\end{align}
for all $i\in i_{-}$ since the assumption $I_{\eta}(x_{i}^{*},y)>0$ implies $q_{\eta}(x_{i}^{*},y)>0$. Therefore
\begin{align}
-w_{i}^{*}\frac{m_{b_{i}}}{Q_{\eta}(\overline{\mathcal{D}}_{M_{b_{i}-1}})} < w_{i}^{*} - \omega_{\eta}(\overline{\mathcal{D}}_{M_{i,J}}) \le 0. \label{eq:Cond3}
\end{align}

Consider 2 cases. Case 1: $b_{i} = O(n)$. Note the assumption Var$[I_{\eta}(x_{i}^{*},y)]<\infty$ implies $Q_{\eta}(\overline{\mathcal{D}}_{M_{j}}) = O_{p}(j)$. Therefore by equation \eqref{eq:Cond3}
\begin{align}
|w_{i}^{*} - \omega_{\eta}(\overline{\mathcal{D}}_{M_{i,b_{i}-1}})| & \le w_{i}^{*}\frac{m_{b_{i}}}{Q_{\eta}(\overline{\mathcal{D}}_{M_{b_{i}-1}})} \\
&= O_{p}(M_{b_{i}-1}^{-1}) \\
&= O_{p}(n^{-1}).
\end{align}

Case 2: $b_{i} = o(n)$. For this case
\begin{align}
\omega_{\eta}(\overline{\mathcal{D}}_{M_{i,J}}) = \frac{q_{\eta}(\overline{\mathcal{D}}_{M_{i,b_{i}}})}{Q_{\eta}(\overline{\mathcal{D}}_{M_{J}})} \le \frac{Q_{\eta}(\overline{\mathcal{D}}_{M_{b_{i}}})}{Q_{\eta}(\overline{\mathcal{D}}_{M_{J}})} = o_{p}(1).
\end{align}
The above implies for case 2
\begin{align}
w_{i}^{*} - \omega_{\eta}(\overline{\mathcal{D}}_{M_{i,J}}) \rightarrow w_{i}^{*} \ge 0
\end{align}
which contradicts equation \eqref{eq:Cond3}; therefore, if $b_{i} = o(n)$ then $i\notin i_{-}$. Thus case 1 and 2 establish for all $i\in i_{-}$
\begin{align}
w_{i}^{*} - \omega_{\eta}(\overline{\mathcal{D}}_{M_{i,b_{i}-1}}) = O_{p}(n^{-1}).
\end{align}
Now for the set $i_{+} = \{i:w'_{i,J+1}>0\}$ we note that
\begin{align}
\sum_{i=1}^{d} \left[ w_{i}^{*} - \omega_{\eta}(\overline{\mathcal{D}}_{n_{i}}) \right] = \sum_{i\in i_{+}} \left[w_{i}^{*} - \omega_{\eta}(\overline{\mathcal{D}}_{n_{i}})\right] - \sum_{i\in i_{-}} \left[w_{i}^{*} - \omega_{\eta}(\overline{\mathcal{D}}_{n_{i}})\right] = 0.
\end{align}
Therefore,
\begin{align}
\sum_{i\in i_{+}} \left[w_{i}^{*} - \omega_{\eta}(\overline{\mathcal{D}}_{n_{i}})\right] = \sum_{i\in i_{-}} \left[w_{i}^{*} - \omega_{\eta}(\overline{\mathcal{D}}_{n_{i}})\right] = O_{p}(n^{-1}),
\end{align}
which concludes the proof of the Lemma. The LOAD design by its definition satisfies the conditions of the Lemma which concludes the proof of the Theorem.

\bibliographystyle{elsarticle-harv}
\bibliography{bibtex_entries}

\end{document}